\documentclass[prd,superscriptaddress,nofootinbib,amsmath,amssymb,aps,11pt]{revtex4-2}

\usepackage{soul}
\usepackage[justification=raggedright,singlelinecheck=false]{caption}
\usepackage{subcaption}
\usepackage{hyperref}
\usepackage{float}
\usepackage[normalem]{ulem}
\usepackage{bm}
\usepackage{amsfonts}
\usepackage{latexsym}
\usepackage[latin1]{inputenc}
\usepackage{tabularx}
\usepackage{array}
\usepackage{palatino}
\usepackage{mathpazo}
\usepackage{mathtools}
\usepackage{graphicx} 
\usepackage{appendix}
\usepackage{booktabs}
\usepackage{dcolumn}
\usepackage[dvipsnames]{xcolor}
\usepackage{multirow}
\usepackage{acro}

\DeclareAcronym{EHT}{short = EHT, long = Event Horizon Telescope Collaboration}
\DeclareAcronym{ngEHT}{short = ngEHT, long = Next Generation Event Horizon Telescope Collaboration}
\DeclareAcronym{GRMHD}{short = GRMHD, long = general relativistic magneto-hydrodynamic}
\DeclareAcronym{ZAMO}{short = ZAMO, long = zero angular momentum observer}
\DeclareAcronym{EVPA}{short = EVPA, long = electric vector position angle}
\DeclareAcronym{ISCO}{short = ISCO, long = innermost stable circular orbit}
\DeclareAcronym{TCOs}{short = TCOs, long = timelike circular orbits}
\DeclareAcronym{LIGO}{short = LIGO, long = Laser Interferometer Gravitational-Wave Observatory}
\DeclareAcronym{SMBH}{short = SMBH, long = supermassive black hole}

\begin{document}
        
\title{Polarized Equatorial Emission around Kerr Black Holes with Synchronized Scalar Hair. I. Direct images}
        
\author{Valentin O. Deliyski}
	\email{valentin.deliyski@phys.uni-sofia.bg}
	\affiliation{Department of Theoretical Physics, Faculty of Physics, Sofia University, Sofia 1164, Bulgaria}

\author{Galin N. Gyulchev}
	\email{gyulchev@phys.uni-sofia.bg}
	\affiliation{Department of Theoretical Physics, Faculty of Physics, Sofia University, Sofia 1164, Bulgaria}
    \affiliation{Space Research and Technology Institute, Bulgarian Academy of Sciences, Acad. G. Bonchev Str., Bl. 1, 1113 Sofia, Bulgaria}
  
\author{Daniela D. Doneva}
\email{daniela.doneva@uv.es}
\affiliation{Departamento de Astronom\'ia y Astrof\'isica, Universitat de Val\`encia,
	Dr. Moliner 50, 46100, Burjassot (Val\`encia), Spain}
\affiliation{Theoretical Astrophysics, Eberhard Karls University of T\"ubingen, 72076 T\"ubingen, Germany}

\author{Petya G. Nedkova}
    \email{pnedkova@phys.uni-sofia.bg}
    \affiliation{Department of Theoretical Physics, Faculty of Physics, Sofia University, Sofia 1164, Bulgaria}

\author{Stoytcho S. Yazadjiev}
	\email{yazad@phys.uni-sofia.bg}	
	\affiliation{Department of Theoretical Physics, Faculty of Physics, Sofia University, Sofia 1164, Bulgaria}
	\affiliation{Institute of Mathematics and Informatics, 	Bulgarian Academy of Sciences, Acad. G. Bonchev Str., Bl. 8, Sofia 1113, Bulgaria}
    
\begin{abstract}

We investigate the polarization properties of the direct images of a geometrically and optically thin accretion disk around fully self-consistent models of rotating Kerr black holes with synchronized bosonic hair. The presence of a massive scalar field alters the geodesic structure of the spacetime and thus leaves an imprint on the polarization of radiation emitted near the black hole horizon. To study this effect, we employ a simple analytical model of a geometrically thin accretion disk, orbiting in the equatorial plane and emitting synchrotron radiation. The main deviation from a corresponding Kerr black hole in general relativity is found to be a dephasing in the twist of the polarization vector, which is surprisingly larger for the least scalarized solutions we consider. This behavior suggests that polarization observables are primarily sensitive to local geometric and transport effects along photon trajectories rather than to the overall scalar field strength. Furthermore, our results demonstrate that while equatorial magnetic fields produce qualitatively similar polarization patterns to Kerr black holes in general relativity, vertical magnetic fields at high observer inclinations can lead to a characteristic reversal of the twist direction of the polarization vector.
\end{abstract}

\maketitle

\section{Introduction}

Recently the \ac{EHT} has produced radio images of the \ac{SMBH} candidates in the cores of the Milky Way \cite{EHT_SgrA_I} and the galaxy M87 \cite{EHT_M87_I} at an unprecedented spatial resolution. These results are remarkable because they offer the first experimental glimpses at the shadow of such extreme compact objects, as well as the highly lensed images of their surrounding emission medium. The main observables from this experiment are the total flux ($\approx 0.5$ mJy for M87* and $\approx 1.5$ mJy for SgrA* at the 230 GHz observational frequency of the \ac{EHT}), the image morphology and variability \cite{EventHorizonTelescope:2019pgp,EventHorizonTelescope:2022exc,2026arXiv260100394B}, and its polarization \cite{EventHorizonTelescope:2021bee,EventHorizonTelescope:2024hpu}. The \ac{EHT} results are most consistent with the surrounding emission medium being an optically thin accretion disk, emitting synchrotron radiation. The magnetic field of the disk is believed to be dynamically important near the horizon, and to have an ordered structure on scales comparable to the gravitational radius of the central object \cite{EHT_M87_VII, EHT_sgrA_VIII}.

\bigskip\par

These observations provide a new opportunity to probe strong-field gravity in the immediate vicinity of black holes, where both spacetime geometry and plasma processes play a central role in shaping the observed signals. The study of the polarized images of accretion disks around compact objects predates the observations by the \ac{EHT} collaboration. The basic theory of polarized emission around black holes in GR was outlined in a range of classical works \cite{Bardeen1972ApJ, Connors1980ApJ, Ipser1982, Chen1991ApJ, Agol1997PhDT}. Subsequent studies on the topic have mainly focused on either the radio-to-millimeter, near infrared or X-ray bands. The focus of this paper is entirely on observational frequencies of the \ac{EHT}, which lie in the radio band. The main emission mechanism at radio to near infrared wavelengths is believed to be optically thin synchrotron radiation, which can be used as a probe for the magnetic field structure, accretion regime and even the properties of the central compact objects  \cite{Quataert_2000, Agol_2000, Broderick2006, Shahzamanian2015}.\bigskip\par

Prompted by the success of the \ac{EHT} observations, the topic of synchrotron emission around supermassive compact objects in the radio band has recently seen a significantly increased interest \cite{Rosales2018, Dexter2020, Himwich2020, Chael_2021, GRAVITY2023,Chael_2023, Gelles_2025, Uniyal2025, EHT_OJ287, Wong_2026, Cruz-Osorio_2026}. One notable development, which resulted from this, is the analytical model presented in \cite{Narayan_2021}. It describes a thin ring of fluid, orbiting on a circular geodesic around a Schwarzschild black hole, and emitting synchrotron radiation. This model was shown to be remarkably well suited to qualitatively reproducing the observed twist of the polarization vector in the 2017 \ac{EHT} observations, and it was later extended to the Kerr metric in \cite{M_Johnson_2021}.\bigskip\par

Since then there have been numerous studies concerning how compact objects, which are not described by the Kerr metric, would appear under the observational conditions of the \ac{EHT}. Many of these studies have focused on the morphological appearance of horizonless compact objects \cite{Kocherlakota2021, EHT_2022_Sgr_A_ehtim, PhysRevD.111.064068, Eichhorn_2023, Saurabh2024} and their polarization properties \cite{Wormhole_paper, PhysRevD.108.104049, Zeng:2026ntc, chen2026}. Other studies on the topic have looked at black holes in various different theoretical frameworks \cite{Qin2023, Gauss_Bonnet_pol, Guo2024, Shi2024, qin2025, Tsanimir_2025}. These studies suggest that while the non-Kerr nature of the central compact object does leave an imprint in the observations, the current generation \ac{EHT} does not have the necessary spatial resolution to definitively rule out a large class of non-Kerr spacetimes. 

\bigskip\par

In the present study, we explore the polarization signature of Kerr black holes with synchronized scalar hair. Despite the growing interest in alternative compact objects, the polarization properties of scalarized Kerr black holes in the EHT regime remain largely unexplored. A remarkable feature of these solutions is that under a synchronization condition, there exist stationary models of massive, time-periodic scalar fields around a central rotating Kerr black hole. They  continuously interpolate between the Kerr solution and rotating boson stars \cite{hairysol1, herdeiro2014,herdeiro2015, collodel2020rotating}, based on the normalized Noether charge $q$ of the solution. Moreover, these are among the limited set of fully consistent beyond-Kerr solutions which can produce observable deviations not only for stellar black holes, but also for supermassive ones relevant for the ETH observations. For $q \ne 0$, the geodesic structure can be drastically different from that of the Kerr black hole \cite{cunha2016shadows, Vincent2016shadows, Cunha2016chaotic_lensing,Collodel_2021, Galin_scalar_shadow, Galin_2026},  which prompts the question of how this modified structure affects the observed polarization of radiation emitted from the vicinity of the black hole horizon. The question itself is highly non-trivial to answer, as the emitted radiation naturally depends on the plasma physics of the emission medium, which is itself coupled to the gravity physics in a non-linear way. Moreover, the polarization of this radiation is affected by the process of parallel transport to the observer. In the models we consider in this paper, the scalar field assumes a toroidal distribution outside the black hole horizon, which modifies the spacetime geometry away from the horizon in a nontrivial way and can thus have a non-negligible effect on the observed polarization. \bigskip\par

For the models we consider, the scalar field forms a toroidal distribution outside the black hole horizon, modifying the spacetime geometry away from the horizon in a nontrivial way and potentially leaving a measurable imprint on the observed polarization. To isolate the role of spacetime geometry, we employ simplified emission models that capture the essential features of synchrotron radiation without relying on full GRMHD simulations. We show that this leads to distinctive signatures in the polarization pattern, most notably in the form of a systematic dephasing of the electric vector position angle, which is particularly sensitive to local deviations in the geodesic structure. The simulations we present in this paper are the first polarized images of accretion disks around fully self-consistent numerical models of black holes with self-gravitating scalar fields.\bigskip\par

This paper is organized as follows. In section \ref{Sec:Solutions} we describe the considered theory, as well as the properties and classification of the employed black hole solutions. In section \ref{Sec:Emission_model}, we describe the theoretical framework in which we model the emission medium around scalarized black holes. Section \ref{Sec:Ray_tracing} provides an overview of our methodology for constructing the polarized images. In section \ref{Sec:Results} we present our results and analysis. The paper ends with conclusions in section \ref{Sec:Conclustions}.

\section{Kerr black holes with synchronized scalar hair}\label{Sec:Solutions}
We consider Einstein gravity minimally coupled to two dynamical scalar fields, $\varphi=(\varphi^{1},\varphi^{2})$. The scalar fields can be interpreted as generalized coordinates on an abstract two-dimensional Riemannian manifold $({\cal E}_2, \gamma_{ab}(\varphi))$, referred to as the target space, which is endowed with a positive-definite metric $\gamma_{ab}(\varphi)$. The action of the theory is given by
\begin{equation}
	\label{TMST action}
	S = \frac{1}{4 \pi G} \int d^4x \sqrt{-g}\left(\frac{R}{4} - \frac{1}{2}g^{\mu \nu}\gamma_{ab}(\varphi)\partial_\mu \varphi^a \partial_\nu \varphi^b - V(\varphi) \right).
\end{equation}
Here $V(\varphi)$ denotes the scalar field potential which we take in the following form
\begin{equation}
	V(\psi) = \frac{1}{2}\mu^2 \psi^2,
\end{equation}
where $\mu$ is the scalar field mass. In the special case of flat target space metric, the model reduces to that of~\cite{herdeiro2014}, describing a single complex scalar field $\Psi=\varphi^1 + i \varphi^2$. \bigskip\par

In order to evade the no-scalar-hair theorems \cite{Heusler1996,Herdeiro:2015waa,Yazadjiev:2025ezx}, the scalar fields must exhibit explicit time dependence. We therefore adopt the ansatz
\begin{equation}
	\label{lucas scalar field ansatz}
	\varphi^1 = \psi(r, \theta)\cos(\omega_s t + m \phi), \quad \varphi^2 = \psi(r, \theta)\sin(\omega_s t +m \phi),
\end{equation}
which is compatible with the circular symmetry of the metric given below in eq. \eqref{lucas line element}. It also ensures that the field equations remain stationary. Here, $\omega_s$ is a real frequency parameter and $m$ is an integer azimuthal harmonic index. Additional details can be found in \cite{collodel2020rotating,herdeiro2015}.\bigskip\par

Stationary black hole solutions can exist if the so-called synchronization condition is respected. It guarantees that there is no scalar flux through the horizon. In explicit form, the synchronization condition requires the angular frequency of the scalar field $\omega_s$ to satisfy $\omega_s=m \Omega_{H}$, where $\Omega_H$ is the angular frequency of the black hole horizon and $m$ is the azimuthal number of the scalar fields. The existence of Kerr black holes with synchronized hair in the perturbative regime (i.e. without taking the back reaction of the scalar fields on the spacetime geometry), the so-called scalar clouds,  was first discovered in \cite{hairysol1}. The fully non-linear and self-consistent numerical solutions describing Kerr black holes with synchronized scalar hair were constructed in \cite{herdeiro2014}. Fully non-linear generalizations of \cite{herdeiro2014} for scalar fields with a non-flat target space being a maximally symmetric 2-dimensional manifold with Gaussian curvature $\kappa$ were numerically constructed in \cite{collodel2020rotating}. For convenience and later use, we give the line element employed in these solutions:
\begin{equation} 
\label{lucas line element}
    ds^2 = -\mathcal{N}e^{2F_0}dt^2 +e^{2F_1}\left(\frac{dr^2}{\mathcal{N}} + r^2 d\theta^2\right) +e^{2F_2}r^2 \sin^2\theta \left(d \phi - \frac{\omega}{r}dt \right)^2,
\end{equation}
where $\mathcal{N} = 1 - \frac{r_\mathrm{H}}{r}$, with $r_\mathrm{H}$ denoting the location of the event horizon in these coordinates. The functions $F_0, F_1, F_2$, and $\omega$ depend only on the variables $r$ and $\theta$. This class of solutions also contains Kerr black holes in GR. For this case the line element \eqref{lucas line element} reduces to the well known Kerr metric, but in a shifted radial coordinate, which is related to the Boyer-Lindquist radial coordinate $R$ by $r=R-a^{2}/R_{\mathrm{H}}$, where $a$ is the spin parameter and $R_H$ is the event horizon radius. \bigskip\par

Kerr black hole solutions with synchronized  hair are in principle characterized by three global conserved charges, namely  the  ADM mass $M$, the total angular momentum $J$, and the Noether charge $Q$. It is convenient to introduce the normalized Noether charge $ q = \frac{mQ}{J} $ as a measure of hairiness. The angular momentum of the scalar field is quantized as $ J_{\psi} = mQ $, which leads to the expression $ q = \frac{J_{\psi}}{J} $. Solutions with $ q \approx 0 $ represent scalar clouds that do not backreact on the metric \cite{hairysol1}, while the pure boson star limit corresponds to $ q \approx 1 $ \cite{doneva1, doneva2}.

\begin{figure}[H]
    \centering
    \includegraphics[width=0.55\textwidth]{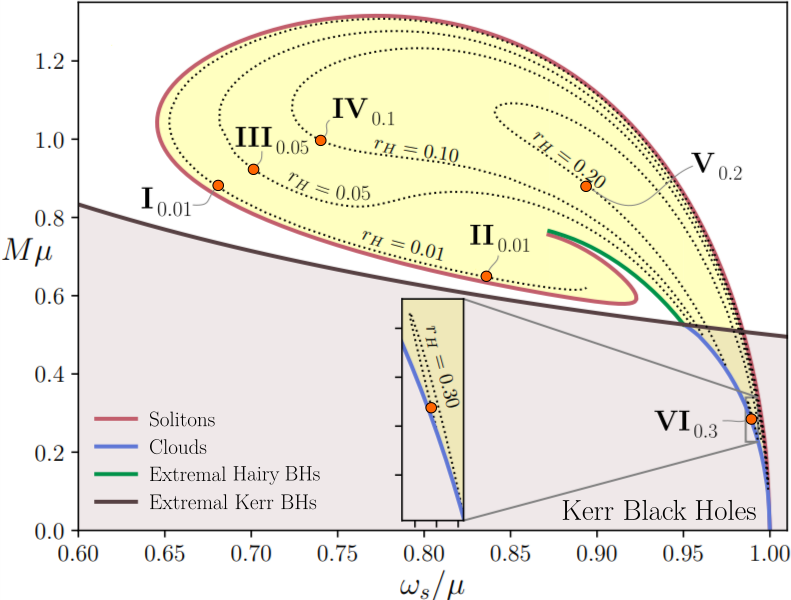}    
    \caption{\small In the $M-\omega_s$ plane, we present solution curves for fixed $r_H$ from \cite{collodel2020rotating} with $\kappa = 0$. Kerr black holes exist below the thick black line which represents extremal solutions, while hairy black holes exist in the yellow region. Orange dots indicate the solutions used for the construction of accretion discs. They are labeled as $\mathbf{X}_{\,v}$, where $\mathbf{X}$ denotes the model number, and $v$ indicates the black hole horizon radius $r_\mathrm{H}$. Further details of these solutions can be found in Appendix \ref{Appendix Solutons}.}\label{fig:M-Omega Space}
\end{figure} 

In the present work, we focus exclusively on models with vanishing target space curvature, i.e. the $\kappa=0$ case in \cite{collodel2020rotating}, which coincides with the original solutions obtained in \cite{herdeiro2014}. Models with non-flat target curvature will be analyzed in a forthcoming paper (for partial results see \cite{GCandHairyKerr}). The domain of existence in the $ M - \omega_s $ plane for hairy black holes with $ m=1 $ is shown in Fig. \ref{fig:M-Omega Space} (see \cite{herdeiro2014} and \cite{collodel2020rotating}). The extremal Kerr limit with $ a=M$, represented by the thick black curve, delineates the region where Kerr black holes exist (grey region below the curve). Hairy black hole solutions occupy the yellow region, bounded by the red line (solitonic limit with $ q=1 $ and $ r_\mathrm{H}=0 $), the green line (extremal hairy black holes with $ q \ne 0 $), and the blue lines corresponding to scalar cloud solutions ($ q=0 $). \bigskip\par

Dotted lines represent sequences of solutions with constant horizon radii, while orange dots indicate specific black hole solutions used in the construction of accretion disks below. A broader region of existence in $ \omega_s $ is observed for black holes with smaller $ r_\mathrm{H} $ values, which approaches the solitonic limit as the horizon radius tends to zero. For each fixed $ r_\mathrm{H} $, the sequence begins at the Minkowski limit ($ \omega_s/\mu=1 $) and ends at the cloud line. 

\section{Modeling polarized synchrotron emission}\label{Sec:Emission_model}

In order to study how the presence of a synchronized scalar hair affects the polarization signatures of the surrounding accretion flow we apply a phenomenological model of magnetized fluid, orbiting in the equatorial plane around the black hole and emitting synchrotron radiation. This model was initially proposed in \cite{Narayan_2021} to analytically study the qualitative polarization signatures of the \ac{EHT} observations of M87*, without the need to perform computationally expensive \ac{GRMHD} simulations of the accretion flow. While they only considered the central compact object to be a Schwarzschild black hole in GR, this model was later extended to the Kerr black hole in GR in \cite{M_Johnson_2021}, to 4D Gauss-Bonnet black holes in \cite{Gauss_Bonnet_pol} and to arbitrary spherically symmetric spacetimes in \cite{Wormhole_paper}. We now present this model, as it applies to a general axially-symmetric spacetime, given by the line element \eqref{lucas line element}. \bigskip\par

\subsection{Emission model outline}

Describing the polarization state of the emitted radiation is most conveniently done in a local orthonormal tetrad $\{e^\mu_{(a)}\}$\footnote{We follow the standard convention in which Greek indices $\{\alpha, \beta, \gamma, \dotsb\}$ without parentheses are acted upon with the spacetime metric $g_{\mu\nu}$, while Latin indices $\{a, b, c, \dotsb\}$ inside parentheses are acted upon with the Minkowski metric $\eta_{(a)(b)}$.}, corresponding to a \ac{ZAMO}\footnote{The name stems from the choosing $e_{(t)}^\mu$ to be equal to the four-velocity of an observer with zero angular momentum.}. The general form of such a tetrad is\footnote{Note the negative sign in the definition of $e_{(\theta)}^\mu$. We adopt the convention that this tetrad vector points in the positive $z$ direction when evaluated on the equator $\theta = \frac{\pi}{2}$.}:
\begin{subequations}\label{Tetrad_definition}
    \begin{equation}
        e_{(t)} = \sqrt{\frac{g_{\phi\phi}}{g_{t\phi}^2 - g_{tt}g_{\phi\phi}}}\left[\partial_t - \frac{g_{t\phi}}{g_{\phi\phi}}\partial_\phi\right] = \frac{e^{-F_0}}{\sqrt{\mathcal{N}}}\left[\partial_t + \frac{\omega}{r}\partial_\phi\right],
    \end{equation}
    \begin{equation}
        e_{(r)} = \frac{1}{\sqrt{g_{rr}}}\partial_r = \sqrt{\mathcal{N}}{e^{-F_1}}\partial_r,
    \end{equation}
    \begin{equation}
        e_{(\theta)} = -\frac{1}{\sqrt{g_{\theta\theta}}}\partial_\theta = -\frac{e^{-F_1}}{r}\partial_\theta,
    \end{equation}
    \begin{equation}
        e_{(\phi)} = \frac{1}{\sqrt{g_{\phi\phi}}} \partial_\phi = \frac{e^{-F_2}}{r\sin\theta}\partial_\phi.
    \end{equation}
\end{subequations}

\noindent They also obey the following orthogonality conditions:
\begin{subequations}
    \begin{equation}
        g_{\mu\nu}e_{(a)}^\mu e_{(b)}^\nu = \eta_{(a)(b)},
    \end{equation}
    \begin{equation}
        \eta^{(a)(b)}e_{(a)}^\mu e_{(b)}^\nu = g^{\mu\nu}.
    \end{equation}
\end{subequations}

\noindent Using this tetrad, we can construct the local rest frame of the emitting particle by first decomposing its four-velocity $u^\mu$ onto $\{e^\mu_{(a)}\}$ in order to construct its boost parameter $\vec{\beta}$:
\begin{equation}
    \vec{\beta} = \frac{1}{u^{(t)}}(u^{(r)}, u^{(\theta)}, u^{(\phi)}), \quad u^{(a)} = \eta^{(a)(b)}e^\mu_{(b)}u_\mu.
\end{equation}

\noindent We then define the local rest frame of the emitting particle by performing a Lorentz boost along $\vec{\beta}$ of the \ac{ZAMO} tetrad \footnote{We adopt the convention that quantities, evaluated in the local rest frame of the emitting particle are labeled with a hat.}:
\begin{equation}
    \hat{e}^\mu_{(a)} = \Lambda_{(a)}^{\quad(b)}e^\mu_{(b)} ,
\end{equation}
while the components of vectors are transformed with the inverse boost $\left(\Lambda_{(a)}^{\quad(b)}\right)^{-1} \equiv \Lambda_{\quad(a)}^{(b)}$: $\hat{V}^{(a)} = \Lambda_{\quad(b)}^{(a)} V^{(b)}$. The explicit form of the boost matrix (labeling $\vec{\beta}\cdot\vec{\beta} \equiv \beta^2$)\footnote{We note that $\vec{\beta}$ is a spatial three-vector in the \ac{ZAMO} frame. Thus its dot product is taken with respect to the Euclidean metric.} is:
\begin{equation}
\Lambda_{\quad(a)}^{(b)}
=
\begin{bmatrix}
     \gamma             & -\gamma\beta_r                                  & -\gamma\beta_\theta                                 & -\gamma\beta_\phi \\
    -\gamma\beta_r      & 1 + (\gamma - 1)\frac{\beta_r^2}{\beta^2}       & (\gamma - 1)\frac{\beta_r\beta_\theta}{\beta^2}     & (\gamma - 1)\frac{\beta_r\beta_\phi}{\beta^2} \\
    -\gamma\beta_\theta & (\gamma - 1)\frac{\beta_\theta\beta_r}{\beta^2} & 1 + (\gamma - 1)\frac{\beta_\theta^2}{\beta^2}      & (\gamma - 1)\frac{\beta_\theta\beta_\phi}{\beta^2}  \\
    -\gamma\beta_\phi   & (\gamma - 1)\frac{\beta_\phi\beta_r}{\beta^2}   & (\gamma - 1)\frac{\beta_\phi\beta_\theta}{\beta^2}  & 1 + (\gamma - 1)\frac{\beta_\phi^2}{\beta^2}
\end{bmatrix} , \quad \gamma = \frac{1}{\sqrt{1 - \beta^2}}
\end{equation}
while its inverse is recovered with the substitution $\vec{\beta}\rightarrow-\vec{\beta}$. We now introduce the magnetic field in the rest frame of the emitting particle $\vec{B} = \left(\hat{B}^{(r)}, \hat{B}^{(\theta)}, \hat{B}^{(\phi)}\right)$, and the local three-momentum of the emitted photon $\vec{k} = \frac{1}{\hat{k}^{(t)}}\left(\hat{k}^{(r)}, \hat{k}^{(\theta)}, \hat{k}^{(\phi)}\right)$. We can use these quantities to compute the emission angle $\xi$:
\begin{equation}
    \sin\xi = \frac{|\vec{k}\times\vec{B}|}{|\vec{B}|},
\end{equation}
where we use the fact that by construction we have $|\vec{k}| = 1$. It is known that synchrotron emission produces a discrete spectrum, which depends strongly on the angle between the local three-velocity of the emitting particle and the magnetic field. A widely known property of such an emission is that in the ultra-relativistic regime $\gamma >> 1$, it is approximately confined to a cone along the local three-velocity of the particle with an opening angle $\approx \frac{1}{\gamma}$. This allows us to identify the angle, governing the emission process, with the angle $\xi$. Furthermore, it has been shown in the classical work \cite{Westfold_1959}, that in the same $\gamma >> 1$ limit, the spectrum tends to a continuum\footnote{This detail is important for the following reason. Consider a particle, emitting at a set of discrete frequencies $\nu_e^n$. To observe this emission, a geodesic must connect the observer to the particle in question. Due to the combination of gravitational and kinematic redshifts, the observer will be able to detect the particle at frequencies $\nu_\text{d}^n = \frac{u^\mu k_\mu|_\text{obs}}{u^\mu k_\mu|_\text{emitter}}\nu_e^n$, which in general will not be equal to the frequency at which they observe $\nu_\text{obs}$. If instead the particle emits with a continuous spectrum, there will always be an emission frequency $\nu_e$, such that $\nu_d = \nu_\text{obs}$.}, whose specific intensity $I_\nu$ has the following angle dependence: 
\begin{subequations}
    \begin{equation}
        I_\nu \propto \sin^{2/3}\xi,\quad h\nu\ll kT,
    \end{equation}
    \begin{equation}
        I_\nu \propto e^{-x^{1/3}},\,\, x \propto \frac{1}{\sin\xi}, \quad h\nu\gg kT,
    \end{equation}
\end{subequations}
where $T$ is the temperature of the emitting particles. The exact dependence on the frequency and temperature can be found in various works that give fitting formulas, which bridge these two regimes \cite{Mahadevan_1996, Marszewski_2021}. The important property of these profiles for us is that the \ac{GRMHD} models of M87* show an approximate angular dependence $\propto \sin^2\xi$ at the 230 GHz observational frequency of \ac{EHT} \cite{Narayan_2021}. The general relationship between the frequency and angular dependence is $I_\nu\propto \nu^{-\alpha_\nu} \rightarrow I_\nu\propto \sin^{1+\alpha_\nu}\xi$. Thus the 2017 M87* observations are well described with $\alpha_\nu = 1$, which we adopt throughout this paper.\bigskip\par

\noindent We can now define the spatial part of the emitted radiation's polarization vector $\vec{f}$ in the emitting particle's rest frame as follows:
\begin{equation}\label{Pol_vec_def}
    \vec{f} = \frac{\vec{k}\times\vec{B}}{|\vec{B}|}.
\end{equation}

\noindent We can also take advantage of gauge freedom to set $\hat{f}^{(t)} = 0$. Given our definition of the emission angle $\xi$, the polarization vector satisfies:
\begin{equation}
    \hat{f}^{(a)}\hat{f}_{(a)} = \sin^2\xi.
\end{equation}

\noindent In accordance with the geometric optics approximation, it then gets parallel transported along a null geodesic to the observer, i.e.
\begin{subequations}\label{System_to_integrate}
    \begin{equation}\label{Geodesic_eqn}
        k^\nu\nabla_\nu k^\mu = 0,
    \end{equation}
    \begin{equation}\label{Parallel_transport_eqn}
        k^\nu\nabla_\nu f^\mu = 0,
    \end{equation}
\end{subequations}

\noindent subject to the condition $k^\mu f_\mu = 0$. Assuming we have a solution to this system of ODEs (which we do numerically as described in section IV), we can write down expressions for the main observable quantities: the (specific) intensity $I_\nu$ and the \ac{EVPA}:
\begin{subequations}
    \begin{equation}\label{Intensity_definition}
        I_\nu \propto \delta^{3 + \alpha_\nu}\ell_p\sin^{1+\alpha_\nu}\xi,
    \end{equation}
    \begin{equation}\label{EVPA_definition}
        \text{EVPA} = \arctan\left(-\frac{f^{(\phi)}_\text{obs}}{f^{(\theta)}_\text{obs}}\right),
    \end{equation}
\end{subequations}

\noindent where $\delta = \frac{u^\mu k_\mu|_\text{obs}}{u^\mu k_\mu|_\text{emitter}}$ is the Doppler factor and $\ell_p = \frac{\hat{k}^{(t)}}{\hat{k}^{(\theta)}}H$ is the projected thickness of the emission medium along the geodesic\footnote{The factor $\ell_p$ arises from the fact that synchrotron emission is specified per unit volume, and is then integrated along the geodesic. For a geometrically thin accretion disk, this can be captured by a simple projection of the disk thickness $H$ onto the geodesic. Throughout this paper we will set the numerical value of $H$ to 1.}. We assume the emitting particle is moving along a circular timelike geodesic in the equatorial plane, co-rotating with the black hole. Its four-velocity $u^\mu$ is then given by:
\begin{subequations}
    \begin{equation}
        u^\mu_\text{emitter} = u^t\left(1, 0, 0, \Omega\right),
    \end{equation}

\noindent where $\Omega$ is the angular velocity and $u^t$ is determined by the normalization condition $u_\mu u^\mu = -1$. Their explicit expressions are given by:
    \begin{equation}
        \Omega=\frac{-\partial_{r}g_{t\phi}+\sqrt{\partial_{r}g_{t\phi}^2-\partial_{r}g_{tt}\partial_{r}g_{\phi\phi}}}{\partial_{r}g_{\phi\phi}},\quad u^t = \frac{1}{\sqrt{-g_{tt} - 2g_{t\phi}\Omega-g_{\phi\phi}\Omega^2}}.
    \end{equation}
\end{subequations}

\subsection{On the spatial dependence of the observed specific intensity}\label{Subsec:Radial_flux_comment}

One of the main benefits of this model is that it captures the phenomenology associated with the apparent twist of the polarization vector, as seen on the observer's screen, without the need to run full \ac{GRMHD} simulations in order to establish the density, temperature and magnetic field profiles of the emission medium. The magnetic field is in fact an input to the model, allowing one to easily probe the effect of various magnetic field geometries on the observed polarization pattern (see \cite{Narayan_2021, M_Johnson_2021, Wormhole_paper, Gauss_Bonnet_pol, PhysRevD.108.104049, Tsanimir_2025}). What the model does not take into account is that due to the very strong temperature dependence of the emission process, most of the observed emission comes from a narrow region around the \ac{ISCO}. The exact apparent shape of this region will vary based on the inclination of the observer (due to the very strong kinematic blue/red shift near the inner regions of the disk at high inclinations).\bigskip\par

The above cited papers have so far focused on the polarization signatures of single orbits, for which (assuming an axi-symmetric emission region) the relative variation in the observed intensity along their apparent images is controlled entirely by the factors in expression \eqref{Intensity_definition}.  In this paper we consider the polarization pattern across the entire accretion disk. Comparing the polarized intensity across different orbit radii of the disk would require us to specify an additional proportionality factor in \eqref{Intensity_definition}, which specifies the intrinsic emissivity of the disk as a function of the emission point. This factor would take into account all the magneto-hydrodynamic phenomenology of the accretion process. These processes are becoming increasingly more accurately simulated for Kerr black holes in GR, but for the highly scalarized black holes which we consider in this work, such simulations have not been performed. It is possible that the phenomenology of the accretion process in the presence of massive synchronized scalar fields differs greatly from that of Kerr black holes in GR (analogously to how the shadows of these objects differ substantially \cite{cunha2016shadows}). We thus refrain from specifying any intrinsic emissivity of the emission medium and when considering the polarized images of the entire accretion disk, we will focus on the \ac{EVPA} pattern. We will compare the relative intensity of images only when considering emission from the neighborhood of a fixed radius. 

\section{Ray Tracing Setup}\label{Sec:Ray_tracing}

In order to construct polarized images of a thin accretion disk we need to solve the ODE system \eqref{System_to_integrate}. We do this in two stages: First we recast the second order geodesic equation \eqref{Geodesic_eqn} into a system of first order equations, derived from the Hamiltonian formalism:
\begin{equation}\label{Hamilton}
    \dot{x}^\mu = \frac{\partial \mathcal{H}}{\partial k_\mu},\quad \dot{k}_\mu = -\frac{\partial\mathcal{H}}{\partial x^\mu},
\end{equation}
where the derivatives on the left hand side are taken with respect to the affine parameter along the geodesic. The Hamiltonian $\mathcal{H}$ for null geodesics takes the form
\begin{equation}
    \mathcal{H} = \frac{1}{2}g^{\mu\nu}k_\mu k_\nu = 0.
\end{equation}

Due to the fact that the underlying spacetime is stationary and axisymmetric, we easily find two conserved quantities along the null geodesics. These are the photon energy $E := -k_t$ and its azimuthal angular momentum $L_z := k_\phi$. We can then decompose the photon momentum $k_\mu$ in the \ac{ZAMO} basis of an observer, by defining the celestial angles $\{\alpha, \beta\}$:
\begin{equation}
    k^{(\phi)} = k^{(t)}\sin\alpha\cos\beta,\quad k^{(r)} = k^{(t)}\cos\alpha\cos\beta,\quad k^{(\theta)} = k^{(t)}\sin\beta.
\end{equation}

\noindent Using the definition of the tetrad \eqref{Tetrad_definition} we can express the components of $k^{(a)}$ in terms of the coordinate momenta $k_\mu$
\begin{subequations}
    \begin{equation}
        k^{(t)} = \zeta E - \gamma L_z, \quad k^{(r)} = \frac{k_r}{\sqrt{g_{rr}}}, \quad k^{(\theta)} = -\frac{k_\theta}{\sqrt{g_{\theta\theta}}}, \quad k^{(\phi)} = \frac{L_z}{\sqrt{g_{\phi\phi}}},
    \end{equation}
    \begin{equation}
        \zeta = \sqrt{\frac{g_{\phi\phi}}{g_{t\phi}^2 - g_{tt}g_{\phi\phi}}}, \quad \gamma =-\frac{g_{t\phi}}{g_{\phi\phi}}\zeta.
    \end{equation}
\end{subequations}

\noindent This allows us to express the photon momenta at the observer in terms of the celestial angles $\{\alpha, \beta\}$:
    \begin{equation}
        E = k^{(t)}\left(\frac{1 + \gamma\sqrt{g_{\phi\phi}}\sin\alpha\cos\beta}{\zeta}\right), \quad
        k_r = k^{(t)}\sqrt{g_{rr}}\cos\alpha\cos\beta,
    \end{equation}
    \begin{equation}
        k_\theta = -k^{(t)}\sqrt{g_{\theta\theta}}\sin\beta, \quad
        k_\phi = k^{(t)}\sqrt{g_{\phi\phi}}\sin\alpha\cos\beta.
    \end{equation}

\noindent Here the overall factor $k^{(t)}$ only rescales the photon energy and does not influence the geodesic. The above expressions are then taken as initial conditions, parametrized by the celestial angles $\{\alpha, \beta\}$. We discretize the field of view for a given observer and integrate Hamilton's equations, \eqref{Hamilton} for each pair of angles, backwards (i.e. for negative values of the affine parameter), until the ray intersects the accretion disk. We then compute an initial polarization vector from eq. \eqref{Pol_vec_def} and numerically integrate the parallel transport equation \eqref{Parallel_transport_eqn} forward along the geodesic, from the emission point to the observer. Each geodesic corresponds to a pixel on his detector (or screen) at Cartesian coordinates $\{x,y\}$, given by\footnote{In some other sources, the tetrad $e_{(\theta)}$ is defined with the opposite sign, which would also change the sign in the definition of $y$ to $ y = r\sin\frac{k^{(\theta)}}{k^{(t)}}\bigg|_{r = r_{\text{obs}}}$.}

\begin{subequations}\label{Screen_coords}
    \begin{equation}
        x = -r\tan\frac{k^{(\phi)}}{k^{(r)}}\bigg|_{r = r_{\text{obs}}}\xrightarrow[r\rightarrow\infty]{}-r\frac{L_z}{(\zeta E - \gamma L_z)\sqrt{g_{\phi\phi}}}\bigg|_{r = r_{\text{obs}}},
    \end{equation}
    \begin{equation}
        y = -r\sin\frac{k^{(\theta)}}{k^{(t)}}\bigg|_{r = r_{\text{obs}}}\xrightarrow[r\rightarrow\infty]{}r\frac{k_{\theta}}{(\zeta E - \gamma L_z)\sqrt{g_{\theta\theta}}}\bigg|_{r = r_{\text{obs}}}.
    \end{equation}
\end{subequations}

\noindent Here we have made use of the fact that the photon four-momentum satisfies 
\begin{equation}
    k^{(a)}e_{(a)} = \frac{e^{-F_0}}{\sqrt{\mathcal{N}}}k^{(t)}\left[\partial_t + \frac{\omega}{r}\partial_\phi\right] + \sqrt{\mathcal{N}}e^{-F_1}k^{(r)}\partial_r - \frac{e^{-F_1}}{r}k^{(\theta)}\partial_\theta + \frac{e^{-F_2}}{r\sin\theta}k^{(\phi)}\partial_\phi
\end{equation}

\noindent and thus have $k^{(a)}e_{(a)}\xrightarrow[r\rightarrow\infty]{}k^{(t)}\partial_t + k^{(r)}\partial_r$. This, along with the normalization condition $k^{(a)}k_{(a)} = 0$, gives $|k^{(t)}| = |k^{(r)}|$ at $r\rightarrow\infty$. For a photon emitted from the accretion disk we necessarily have that at the asymptotic observer $k{(r)}>0$. This implies $k^{(t)} = k^{(r)}$ at $r\rightarrow\infty$, from which we recover the asymptotic form of \ref{Screen_coords}. On this screen we can define the projection of the transported polarization vector $\{f_x, f_y\}$:
\begin{equation}
    f_x = e_\mu^{(\phi)}f^\mu\big|_{r = r_{\text{obs}}} = f^{(\phi)}\big|_{r = r_{\text{obs}}}, \quad f_y = e_\mu^{(\theta)}f^\mu\big|_{r = r_{\text{obs}}} = f^{(\theta)}\big|_{r = r_{\text{obs}}}
\end{equation}

Throughout this paper we will be comparing the \ac{EVPA} of different such vectors, projected on the observer's screen. It is therefore useful to introduce the signed \ac{EVPA} difference $\Delta\text{EVPA}$, between two such vectors $\vec{f}$ and $\vec{f}^\prime$:
\begin{equation}
    \Delta\text{EVPA} = \arctan\left(\frac{f_xf^\prime_y-f_yf^\prime_x}{f_xf^\prime_x + f_yf^\prime_y}\right).
\end{equation}

Once the system \eqref{System_to_integrate} is numerically solved, we color each pixel based on that ray's source intensity as per \eqref{Intensity_definition} for $\alpha_\nu = 1$. We take the source to be part of a geometrically and optically thin prograde accretion disk, starting at the \ac{ISCO}, whose particles travel on stable \ac{TCOs} in the equatorial plane. Note that one of the considered scalarized models, \textbf{I}$_{0.01}$, has multiple disconnected stability zones for its \ac{TCOs}. In such cases, where multiple \ac{ISCO} exist, we take the disk to start from the outermost one, similar to previous analysis \cite{Galin_scalar_shadow,Galin_2026}. Finally, we take the observer's radial coordinate to be $r_\text{obs} = 10^4 M_\text{ADM}$ for all simulations.\bigskip\par
\newpage
\section{Polarized direct images}\label{Sec:Results}

We consider the six hairy black hole numerical solutions denoted in figure \ref{fig:M-Omega Space}. They were originally obtained in \cite{collodel2020rotating}, and the structure of their \ac{TCOs} was recently studied in \cite{Galin_2026}. Their physical properties are summarized in table \ref{tab_sol_properties}. We will compare each of them with a Kerr black hole in GR with the same ADM mass, and a spin parameter such that it has the same horizon radius. We call this Kerr black hole the  ``Kerr analog'' of the scalarized model. The connection between the horizon radius, given in the coordinates of the line element \eqref{lucas line element}, which we denote by  $r_H$, and the standard Boyer-Lindquist coordinates $R_H$ for a pure Kerr black hole, is given in \cite{herdeiro2015}:
\begin{equation}
    r_H = R_H - \frac{a^2}{R_H},\quad R_H = M_{\text{ADM}} + \sqrt{M_{\text{ADM}}^2 - a^2}.
\end{equation}

\noindent We invert this expression to obtain the desired spin parameter $a_{\text{Kerr}}$ for the Kerr analog in terms of the parameters of the scalarized models \footnote{All the scalarized solutions we consider have $J_\text{ADM} > 0$. We therefore choose the positive root in \eqref{Kerr_spin}.}:
\begin{equation}\label{Kerr_spin}
    a_{\text{Kerr}} = \pm\sqrt{M_{\text{ADM}}^2-\frac{r_H^2}{4}} .
\end{equation}

For all six such models, we analyze the polarization properties of the direct images of a geometrically and optically thin accretion disk for the following observer inclinations and magnetic fields:
\begin{equation*}
    i = 17^\circ , \quad i = 45^\circ , \quad i = 70^\circ
\end{equation*}
\begin{equation*}
    \vec{B} = (0.5, 0, 0.87),\quad\vec{B} = (0.71, 0, 0.71),\quad\vec{B} = (0.87, 0, 0.5),\quad\vec{B} = (0, 1, 0).
\end{equation*}

The magnetic fields are the ones originally considered in \cite{Narayan_2021}, and later used in subsequent studies \cite{M_Johnson_2021, Gauss_Bonnet_pol, Wormhole_paper, PhysRevD.108.104049, Tsanimir_2025}. We note, however, that these fields were originally chosen to be orientated antiparallel to the velocity of the fluid (which was considered to have a negative radial component). Given the drastically different structure of the \ac{TCOs} of the considered scalarized spacetimes \cite{Collodel_2021, Galin_2026}, we choose to limit ourselves to pure Keplerian motion. The inclinations are chosen to cover all possible observation conditions, with the $ i = 17^\circ$ case in particular chosen to represent the galactic target M87* \cite{Akiyama_2019}. \bigskip

We summarize the values for $a_\text{Kerr}$ for each considered model in table \ref{tab:Kerr_analog_spins}. Given the large number of simulations, presenting all of them in detail would unnecessarily encumber this paper. We have therefore chosen to focus on the two most representative models from table \ref{tab_sol_properties} -- these are the highly scalarized model \textbf{II}$_{0.01}$, and the mildly scalarized one \textbf{V}$_{0.2}$. For the rest we present some summarized results in sections \ref{Subsec::deviation_vert_intensity}, \ref{Subsec::deviation_eq_intensity}, \ref{Subsec::deviation_vert_evpa} and \ref{Subsec::deviation_eq_evpa}. \bigskip\par

While our chosen combinations of fluid velocity, magnetic field and observer inclinations do not match those in \cite{M_Johnson_2021}, we have nevertheless used this work as a benchmark of our numerical procedure and found excellent agreement with their analytically driven results. Our code reproduces their polarization patterns, and preserves the value of the complex Penrose-Walker constant (given by eq. (12) in \cite{M_Johnson_2021}), to a relative accuracy of $\approx 10^{-5}$ during parallel transport.

\subsection{Vertical magnetic fields at low inclinations}

We begin our analysis by considering vertical magnetic fields $\vec{B} = (0, 1, 0)$. One notable effect that is observed for such fields is that the intensity takes on a higher value on the receding side of the disk for low inclinations \cite{Narayan_2021} -- an effect attributed to relativistic aberration. Note that this effect is generally quoted as an increase in the intensity ``on the left side of the image'' \cite{Narayan_2021}. This is assuming that the fluid rotates in a clockwise motion when projected on the observer's screen. In our models we have the opposite setup -- the fluid rotates counterclockwise and thus we have an overall reflection of the observed intensity across the observer's $y$-axis. We find that this phenomenological behavior is not altered for orbits sufficiently away from the \ac{ISCO} for any of the considered scalarized models. Below, we present our representative selection of results for the more observationally relevant case -- that of a $i = 17^\circ$ inclination. \bigskip\par

\begin{figure}[H]
    \centering
    \includegraphics[width=1\linewidth]{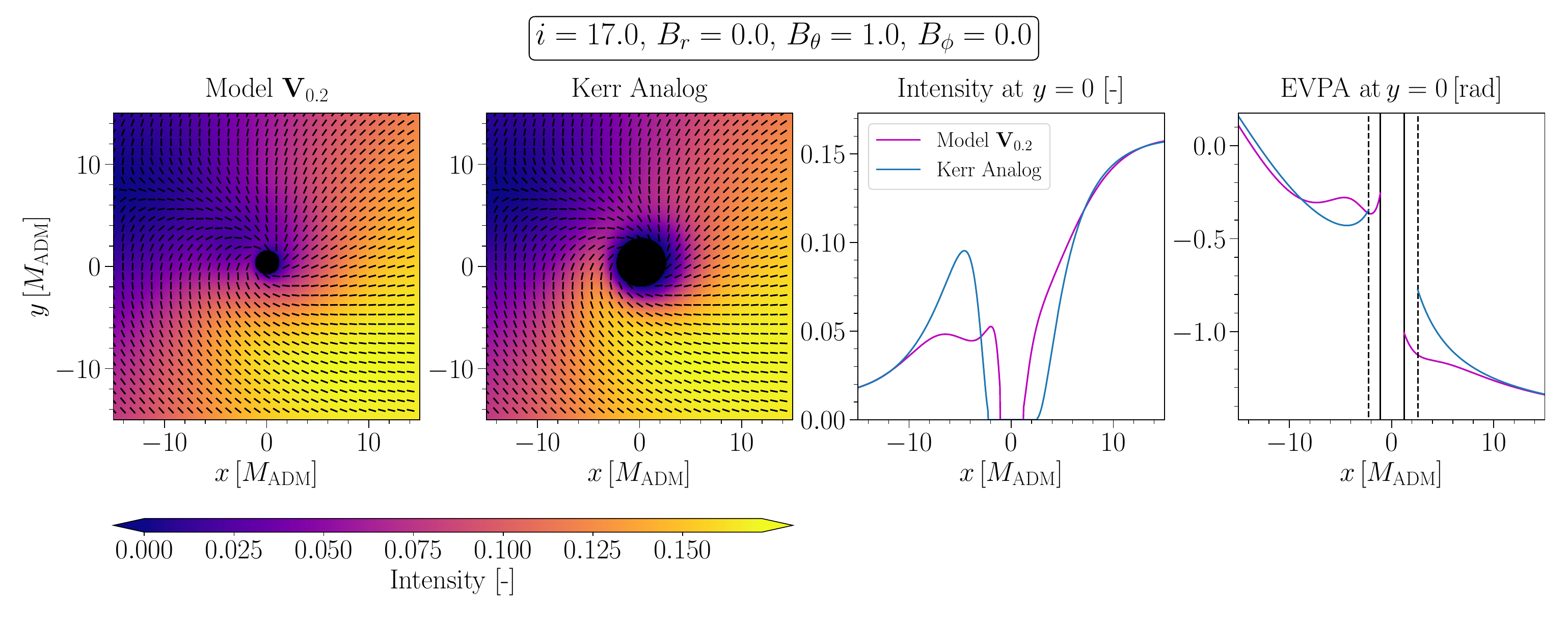}
    \includegraphics[width=1\linewidth]{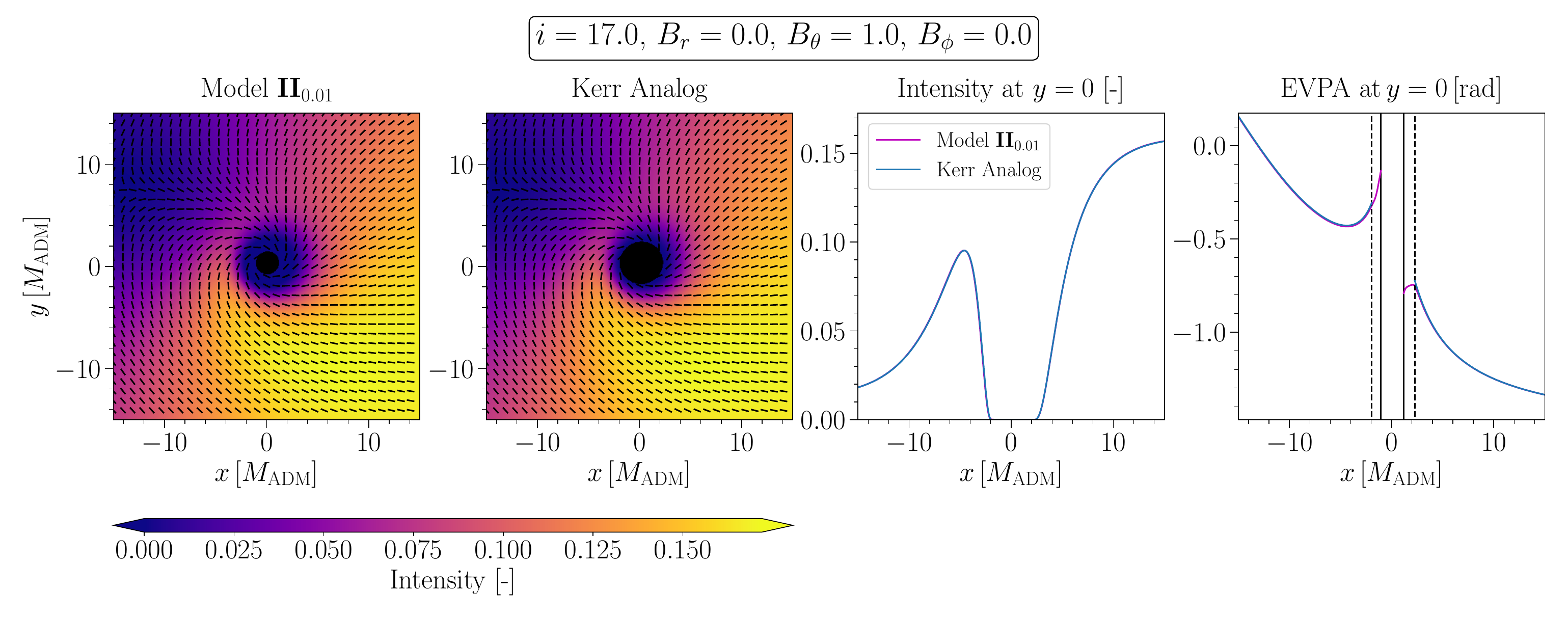}
    \caption{\footnotesize Polarized direct image of a thin accretion disk around the numerical models $\mathbf{II}_{0.01}$, $\mathbf{V}_{0.2}$ and their respective Kerr analogs for a vertical magnetic field, viewed at a $17^\circ$ inclination. The two leftmost panels show the overall polarization pattern and its intensity. The third panel shows the intensity profile as per \eqref{Intensity_definition} at the slice $y = 0$, while the rightmost one shows the \ac{EVPA} on the same $y = 0$ slice. The vertical solid/dashed black lines correspond to the apparent inner edge of the numerical/Kerr solutions.}
    \label{fig:Whole_disk_vert_field_II_V_17_deg}
\end{figure}

The two left panels of figure \ref{fig:Whole_disk_vert_field_II_V_17_deg} show the large scale polarization pattern across the disk in models \textbf{II}$_{0.01}$ and \textbf{V}$_{0.2}$, as well as their Kerr analogs. Note that to better visualize this, we normalize the size of the polarization ticks to unity, and encode the specific intensity at every point with a colormap. The third panel shows the variation of the specific intensity across the $y = 0$ slice of the left two panels. The rightmost panel analogously shows the variation of the \ac{EVPA}.\bigskip\par

We note that on some figures in this paper, the \ac{EVPA} has discontinuities. This is an artifact of the definition \ref{EVPA_definition}. Whenever the polarization vector's $f_y$ component passes through zero, the argument of the $\arctan$ function will jump from $\pm\infty$ to $\mp\infty$, which results in the \ac{EVPA} jumping from $\pm\frac{\pi}{2}$ to $\mp\frac{\pi}{2}$. The actual angular position of the polarization vector on the observer's screen does not experience discontinuities.\bigskip\par

\begin{figure}[H]
    \centering
    \includegraphics[width=1\linewidth]{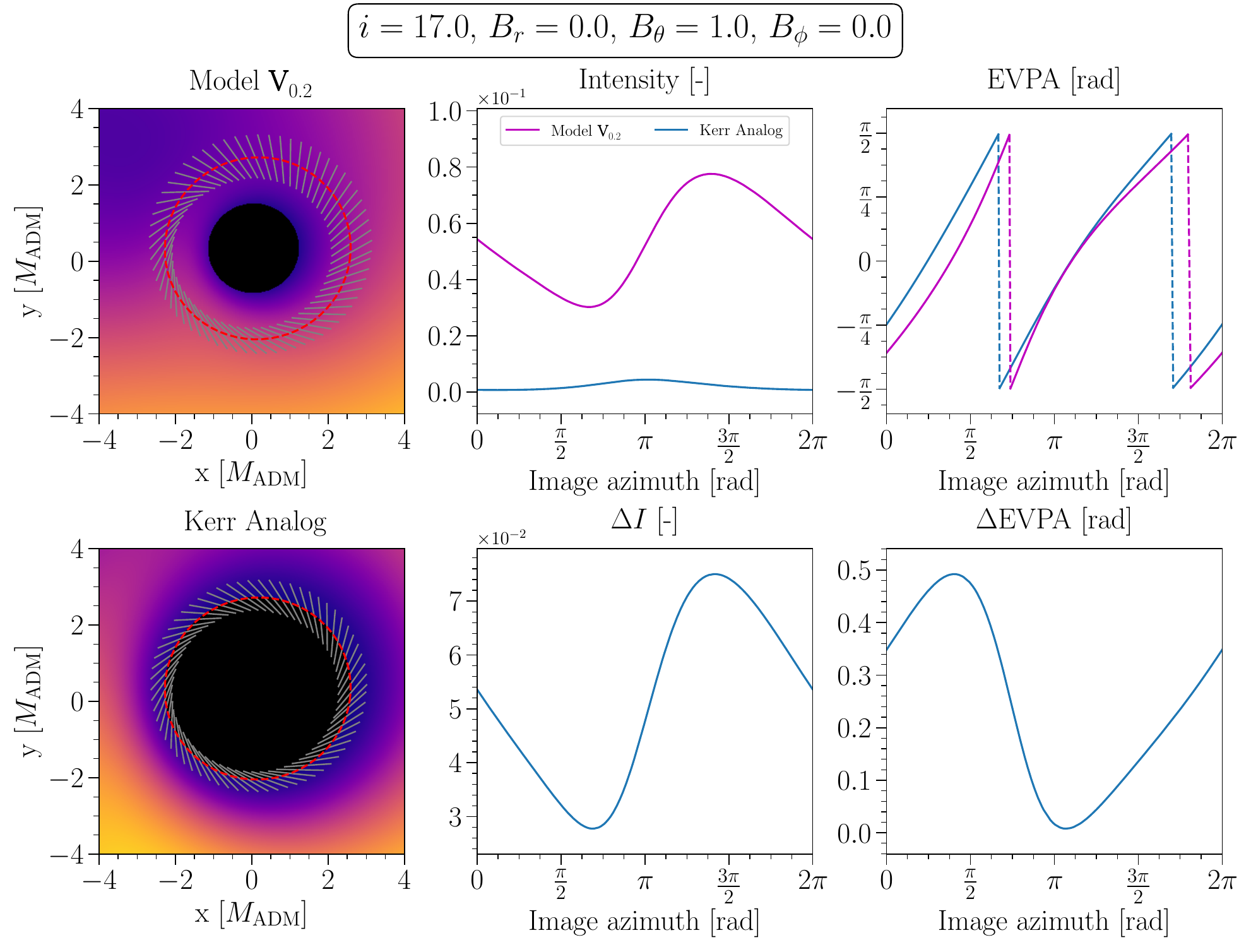}
    \caption{\footnotesize Zoom into the inner regions of the disk from figure \ref{fig:Whole_disk_vert_field_II_V_17_deg}. The leftmost panels show the polarization pattern across the apparent image of the Kerr analog \ac{ISCO}, shown with a red dashed line. The top middle panel shows the intensity variation across this apparent image, going anti-clockwise and starting from the positive $x$-axis. The bottom middle panel shows the difference $\Delta I = I_{\mathbf{V}_{0.2}} - I_{\text{Kerr}}$. The right most panels are analogous, but for the \ac{EVPA}.}
    \label{fig:ISCO_zoom_vert_field_V_17_deg}
\end{figure}

\begin{figure}[H]
    \centering
    \includegraphics[width=1\linewidth]{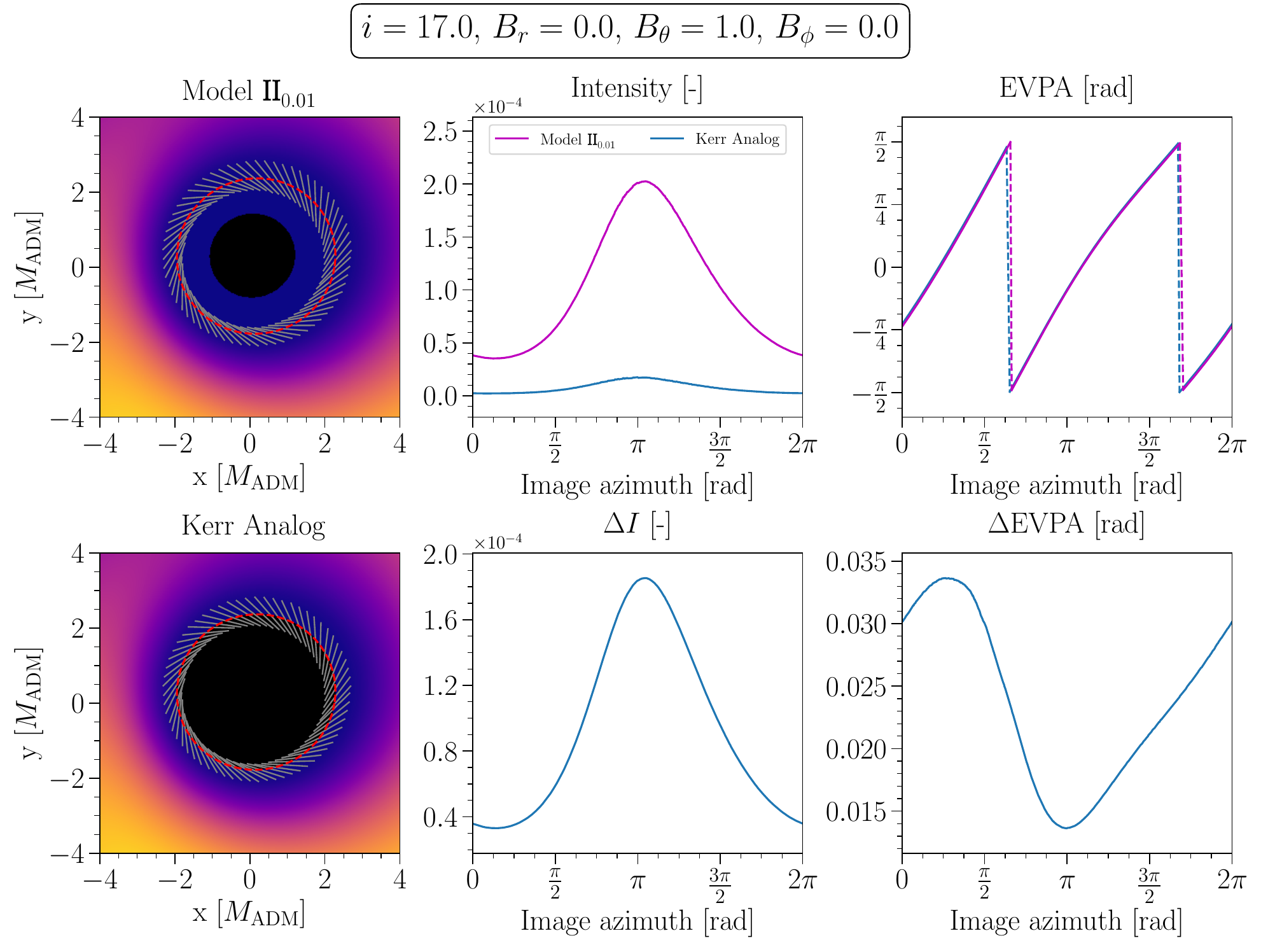}
    \caption{\footnotesize  Zoom into the inner regions of the disk from figure \ref{fig:Whole_disk_vert_field_II_V_17_deg}. The leftmost panels show the polarization pattern across the apparent image of the Kerr analog \ac{ISCO}, shown with a red dashed line. The top middle panel shows the intensity variation across this apparent image, going anti-clockwise and starting from the positive $x$-axis. The bottom middle panel shows the difference $\Delta I = I_{\mathbf{II}_{0.01}} - I_{\text{Kerr}}$. The right most panels are analogous, but for the \ac{EVPA}.}
    \label{fig:ISCO_zoom_vert_field_II_17_deg}
\end{figure}

Figures \ref{fig:ISCO_zoom_vert_field_V_17_deg} and \ref{fig:ISCO_zoom_vert_field_II_17_deg} show a zoom of the inner regions of the disks from figure \ref{fig:Whole_disk_vert_field_II_V_17_deg}. On the leftmost top and bottom panel we plot the polarization pattern across the apparent image of the Kerr analog \ac{ISCO} orbit. The size of the polarization ticks are again normalized for better visualization of the twist pattern, while the intensity is shown with a colormap. On the top middle and right panels we plot the variation of the specific intensity and \ac{EVPA} across these orbits as a function of the azimuthal angle along the image. We take this angle as starting from the positive $x$-axis and increasing in the counterclockwise direction. The bottom middle and rightmost panels then show the difference between the specific intensity and \ac{EVPA} between the scalarized model and its Kerr analog. In all figures the values of the specific intensity (as computed from \ref{Intensity_definition}) are given in arbitrary units.\bigskip\par

Firstly, we notice from the two left panels of figure \ref{fig:Whole_disk_vert_field_II_V_17_deg} that the overall structure of the \ac{EVPA} pattern is qualitatively similar between the numerical models and their Kerr analogs. From the rightmost panels we can see that the deviation between the two, unsurprisingly, grows towards the \ac{ISCO}, but is surprisingly larger for the two least scalarized models $\mathbf{V}_{0.2}$ (figure \ref{fig:Whole_disk_vert_field_II_V_17_deg}) and $\mathbf{VI}_{0.3}$ (not shown). This is completely contrary to what is observed for the shadows, where the ones with a higher normalized Noether charge $q$ show a much higher distortion of the images \cite{cunha2016shadows, Galin_scalar_shadow}.\bigskip\par

This counterintuitive behavior suggests that the polarization signal is not solely governed by the total amount of scalar hair, but rather by how the scalar field modifies the local spacetime structure in the emission region. In particular, even models with small normalized Noether charge $q$ can induce significant changes in the mapping between emission points and image-plane coordinates, as well as in the accumulated phase of the polarization vector through parallel transport. We will look at this effect in more detail in section \ref{EVPA dephasing origin}.

To correctly interpret the third panels on figure \ref{fig:Whole_disk_vert_field_II_V_17_deg}, we note that for low inclinations the images of individual orbits are approximately circular. This means that when looking at the intensity slices through $y = 0$, we can compare values at celestial coordinates $x$ and $-x$, as they will correspond to approximately the same emission radius and thus their relative intensity will depend only on the quantities, appearing in eq. \eqref{Intensity_definition}. We see that for emission radii very close to the \ac{ISCO}, we have the expected result that the receding part of the disk's apparent intensity is lower, but the effect gets quickly reversed as the emission radius increases. We can also see that the approximate location of the maximum intensity for orbits further away from the \ac{ISCO} is on the lower part of the image -- an effect already seen for static black holes and wormholes \cite{Wormhole_paper, Narayan_2021}. It is again notable that the less scalarized solution $\mathbf{V}_{0.2}$ shows larger deviations in the intensity slices on figure \ref{fig:Whole_disk_vert_field_II_V_17_deg}. This non-intuitive result also holds for the other mildly scalarized model in table \ref{tab_sol_properties}, namely $\mathbf{VI}_{0.3}$.\bigskip\par

We now turn our attention to figures \ref{fig:ISCO_zoom_vert_field_V_17_deg} and \ref{fig:ISCO_zoom_vert_field_II_17_deg}, where we have plotted the polarization pattern across the apparent position of the \ac{ISCO} orbit of the corresponding Kerr analog. We see that for model $\mathbf{V}_{0.2}$, the receding intensity increase is visible on the top middle panel.

 In contrast, we find that for model $\mathbf{II}_{0.01}$, the maximum intensity value over the image lies on the approaching side of the disk. We also notice that the intensity computed using eq. \eqref{Intensity_definition} over the apparent position of the Kerr analog \ac{ISCO} is significantly higher for the scalarized models. This is in part due to the fact that for a fixed ADM mass and horizon radius, the focusing effect for the direct images is weaker for the scalarized models, and thus the apparent position of the Kerr analog \ac{ISCO} corresponds to larger radii, where the Doppler factor $\delta^{3 + \alpha_\nu}$ is numerically larger. This weaker focusing effect can be explained by the fact that a significant portion of the ADM mass of the scalarized models is due to the scalar field, which is distributed over a torus around the black hole, rather than being concentrated inside the horizon. An additional contribution to the large intensity difference comes from the fact the Kerr analogs are near extremal (see table \ref{tab:Kerr_analog_spins}), which puts their \ac{ISCO} very close to the infinite redshift surface. This effect will be expanded more on in subsections \ref{Subsec::deviation_vert_intensity} and \ref{Subsec::deviation_eq_intensity}.\bigskip\par

An interesting feature of model $\mathbf{II}_{0.01}$, as compared to $\mathbf{V}_{0.2}$, is the larger central intensity depression. This is best seen on figure \ref{fig:Whole_disk_vert_field_II_V_17_deg}, but also visible on figures \ref{fig:ISCO_zoom_vert_field_V_17_deg} and \ref{fig:ISCO_zoom_vert_field_II_17_deg} as a difference in the intensity across the Kerr analog \ac{ISCO} image of almost two orders of magnitude. One needs to be careful comparing these figures across models. The Kerr analogs are all near extremal and the small differences in spin parameters produce a large relative difference in the \ac{ISCO} radii. This makes direct comparisons of the intensity values misleading (even before remembering that the emission model we employ does not specify a radial distribution).\bigskip\par

The rightmost panels of figures \ref{fig:ISCO_zoom_vert_field_V_17_deg} and \ref{fig:ISCO_zoom_vert_field_II_17_deg} show that while the \ac{EVPA} behaves qualitatively similar across the image, it is dephased relative to the Kerr analog. Analogously to figure \ref{fig:Whole_disk_vert_field_II_V_17_deg}, this dephasing is larger for the less scalarized model $\mathbf{V}_{0.2}$. We also observe this surprising pattern across all six models in table \ref{tab_sol_properties}. The largest deviations from their respective Kerr analogs are displayed by the models $\mathbf{V}_{0.2}$ and $\mathbf{VI}_{0.3}$, which have the smallest normalized Noether charge $q$.

\subsection{Vertical magnetic fields at high inclination}

\begin{figure}[H]
    \centering
    \includegraphics[width=0.95\linewidth]{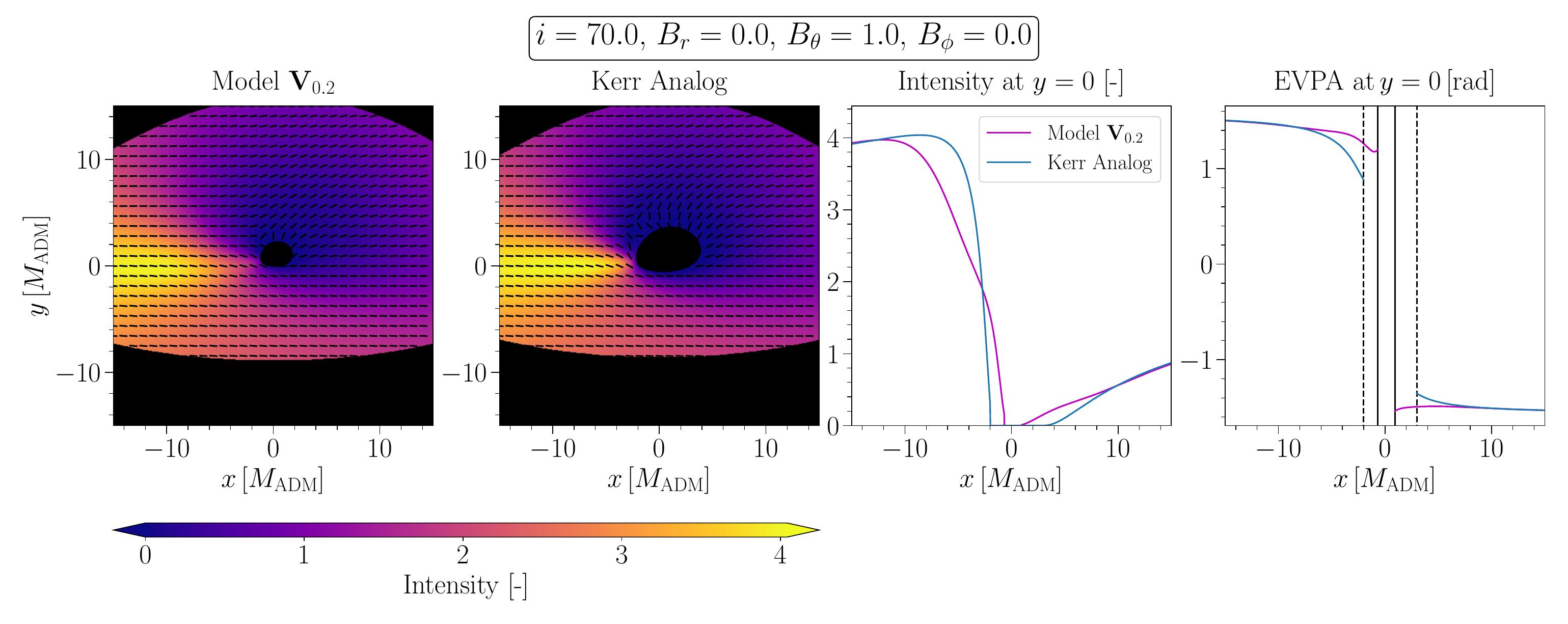}
    \includegraphics[width=0.95\linewidth]{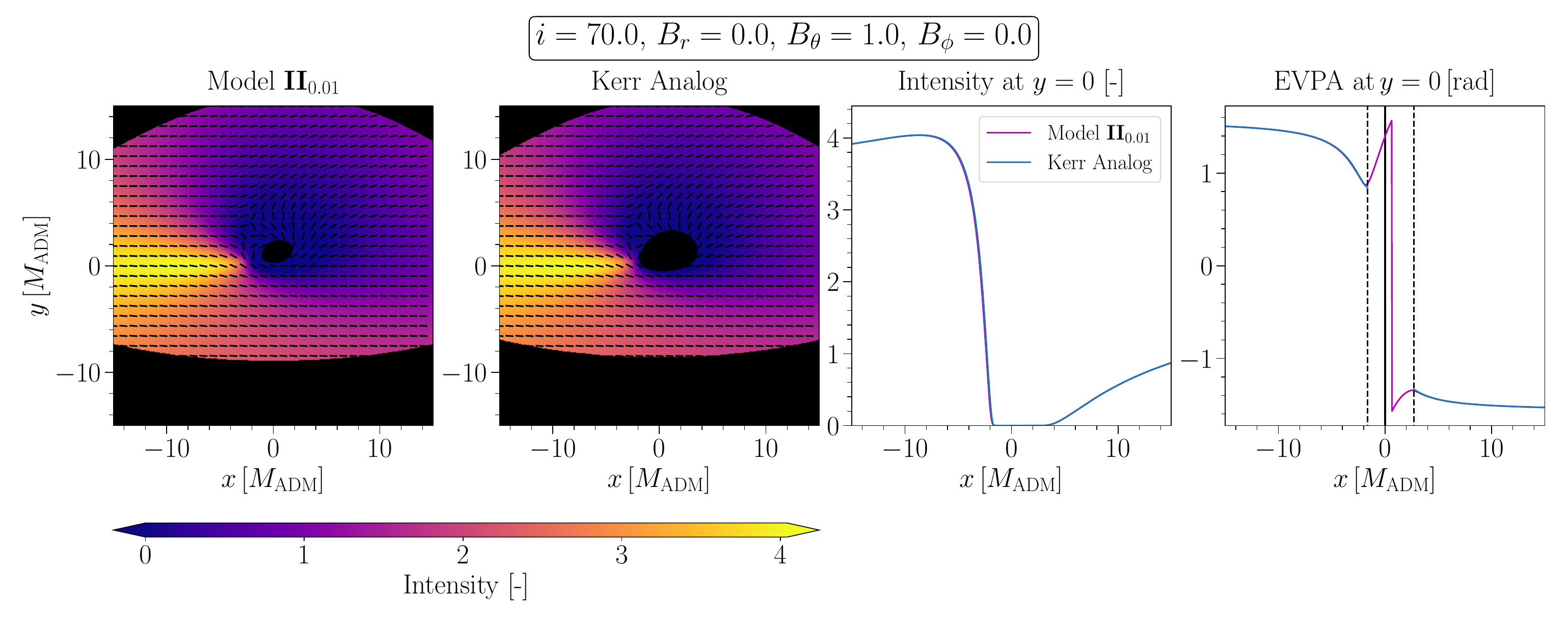}
    \caption{\footnotesize Polarized direct image of a thin accretion disk around the numerical models $\mathbf{II}_{0.01}$, $\mathbf{V}_{0.2}$ and their respective Kerr analogs for a vertical magnetic field, viewed at a $70^\circ$ inclination. The two leftmost panels show the overall polarization pattern and its intensity. The third panel shows the intensity profile as per eq. \eqref{Intensity_definition} at the slice $y = 0$, while the rightmost one shows the \ac{EVPA} on the same $y = 0$ slice. The vertical solid/dashed black lines correspond to the apparent inner edge of the numerical/Kerr solutions.}
    \label{fig:Whole_disk_vert_field_II_V_70_deg}
\end{figure}

We continue our analysis of vertical fields with the high inclination $i = 70^\circ$ case. The overall polarization pattern across the disk is presented on the two leftmost panels of figure \ref{fig:Whole_disk_vert_field_II_V_70_deg}, along with intensity and \ac{EVPA} slices for $y = 0$ on the image. Looking at the intensity slices we see that at higher inclinations the relativistic aberration, leading to a higher intensity on the receding side of the disk, is overcompensated by a strong Doppler boost on the approaching side. This effect is also observed for the lower inclination of $45^\circ$. Interesting to note is that the intensity slice of the highly scalarized model $\mathbf{II}_{0.01}$ is essentially identical to that of its Kerr analog. This was also observed for the low inclination $i = 17^\circ$ case. In contrast, that of the less scalarized model $\mathbf{V}_{0.2}$ shows a more notable quantitative (but not qualitative) difference.

\begin{figure}[H]
    \centering
    \includegraphics[width=1\linewidth]{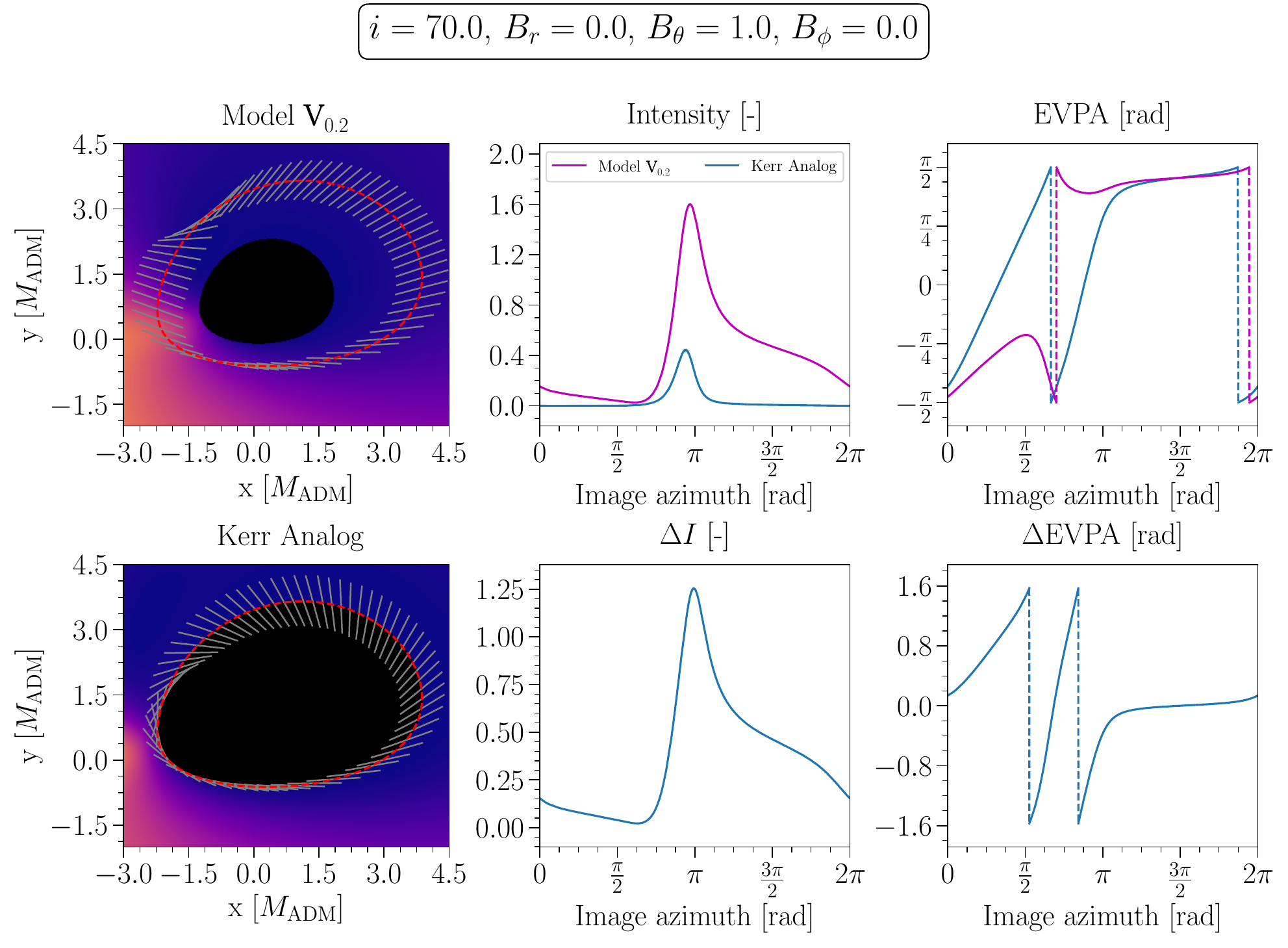}
    \caption{\footnotesize Zoom into the inner regions of the disk from figure \ref{fig:Whole_disk_vert_field_II_V_70_deg}. The leftmost panels show the polarization pattern across the apparent image of the Kerr analog \ac{ISCO}, shown with a red dashed line. The top middle panel shows the intensity variation across this apparent image, going anti-clockwise and starting from the positive $x-$axis. The bottom middle panel show the difference $\Delta I = I_{\mathbf{V}_{0.2}} - I_{\text{Kerr}}$. The right most panels are analogous, but for the \ac{EVPA}.}
    \label{fig:ISCO_zoom_vert_field_V_70_deg}
\end{figure}

\begin{figure}[H]
    \centering
    \includegraphics[width=0.38\linewidth]{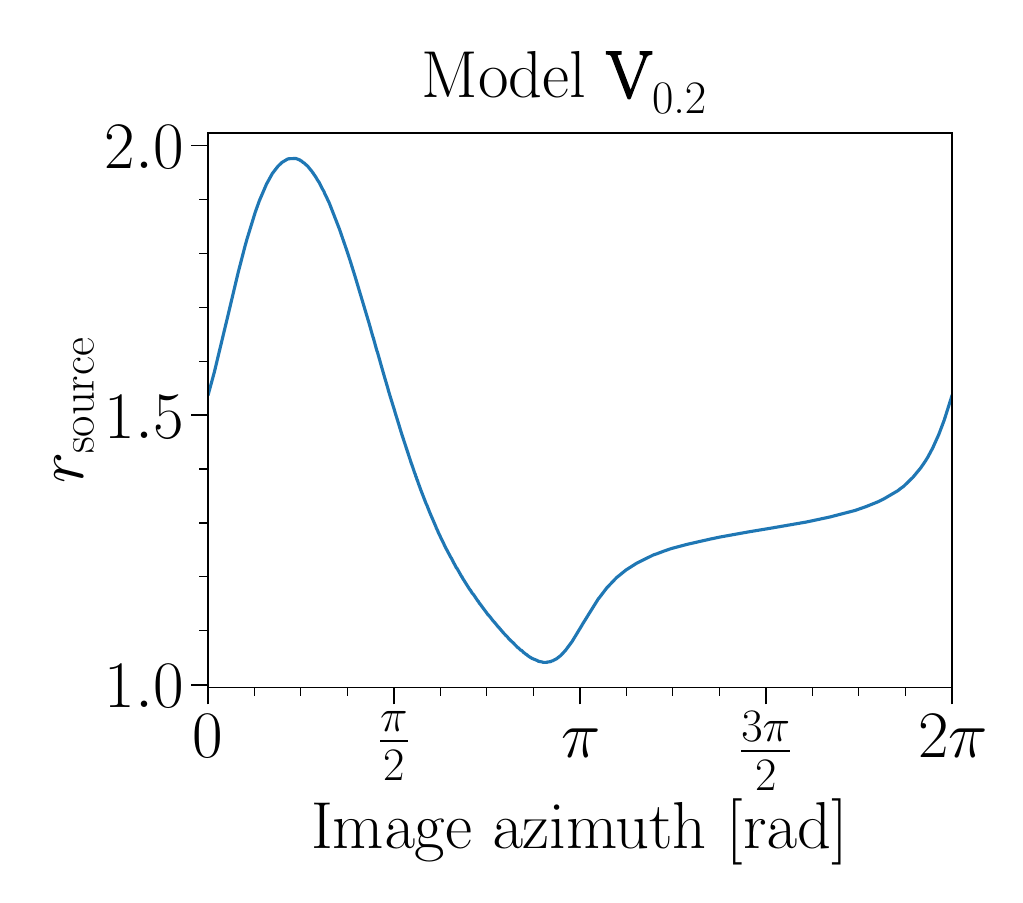}
    \includegraphics[width=0.348\linewidth]{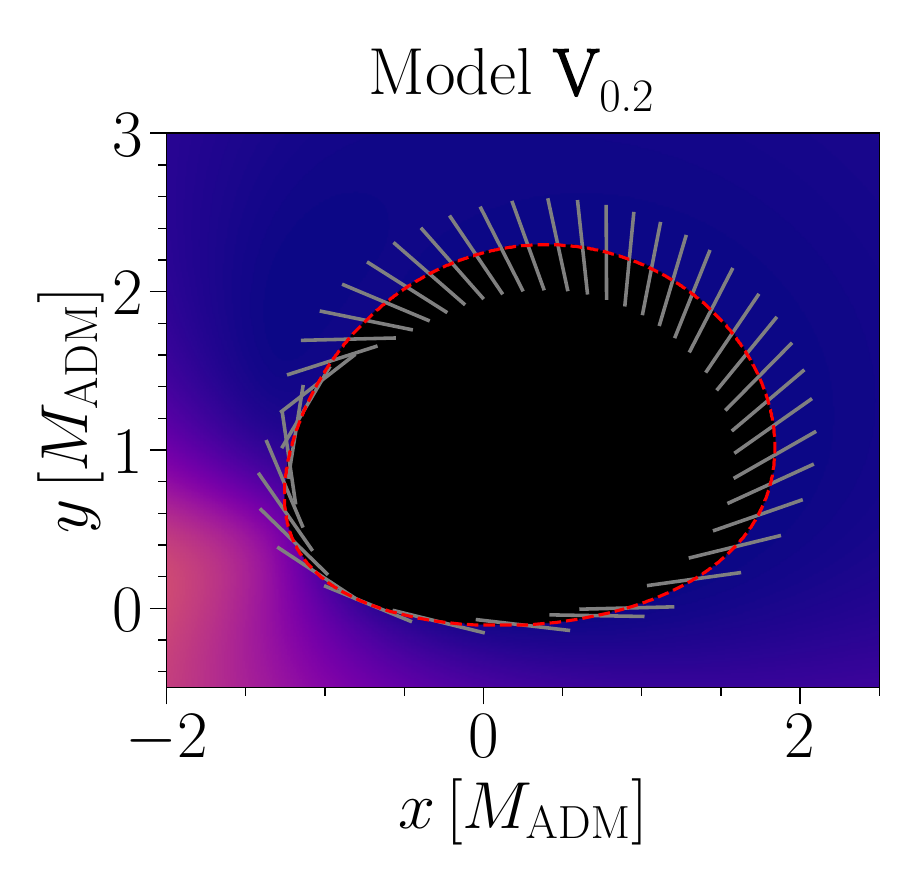}
    \caption{\footnotesize The left panel depicts the variation of the radial source coordinate (in model $\mathbf{V}_{0.2}$) along the image shown in figure \ref{fig:ISCO_zoom_vert_field_V_70_deg}. The right panel shows the polarization pattern across the image of the numerical \ac{ISCO} of model $\mathbf{V}_{0.2}$, rather than that of the Kerr analog.}
    \label{fig:ISCO_zoom_vert_field_V_70_deg_2}
\end{figure}

 We notice that (similar to the low inclination case) in the outer regions of the disk the \ac{EVPA} pattern remains qualitatively similar between the scalarized models 
 and their Kerr analogs. In the inner region though, the leftmost and rightmost panels of figure \ref{fig:ISCO_zoom_vert_field_V_70_deg} show a significant morphological difference for model $\mathbf{V}_{0.2}$, which is present for model $\mathbf{VI}_{0.3}$ as well. The EVPA twist reverses direction compared to the Kerr analog in the upper left portion of the image. We observe this effect only for a vertical magnetic field at high inclinations. It can partially be attributed to the fact that at high inclinations the emission radii, which correspond to the apparent image of the Kerr analog \ac{ISCO}, vary substantially across the image for the considered scalarized models. This is demonstrated in the left panel of figure \ref{fig:ISCO_zoom_vert_field_V_70_deg_2} for model $\mathbf{V}_{0.2}$. Note that the plotted radial coordinate is the one from line element \eqref{lucas line element}. In these coordinates the position of the Kerr analog \ac{ISCO} is $r^\text{ISCO}_\text{Kerr} = 0.4334$. We see that the radial source coordinates along the images are substantially larger than $r^\text{ISCO}_\text{Kerr}$, with a large variation. This again shows that for a fixed ADM mass, the focusing effect is weaker. 

\begin{figure}[H]
    \centering
    \includegraphics[width=1\linewidth]{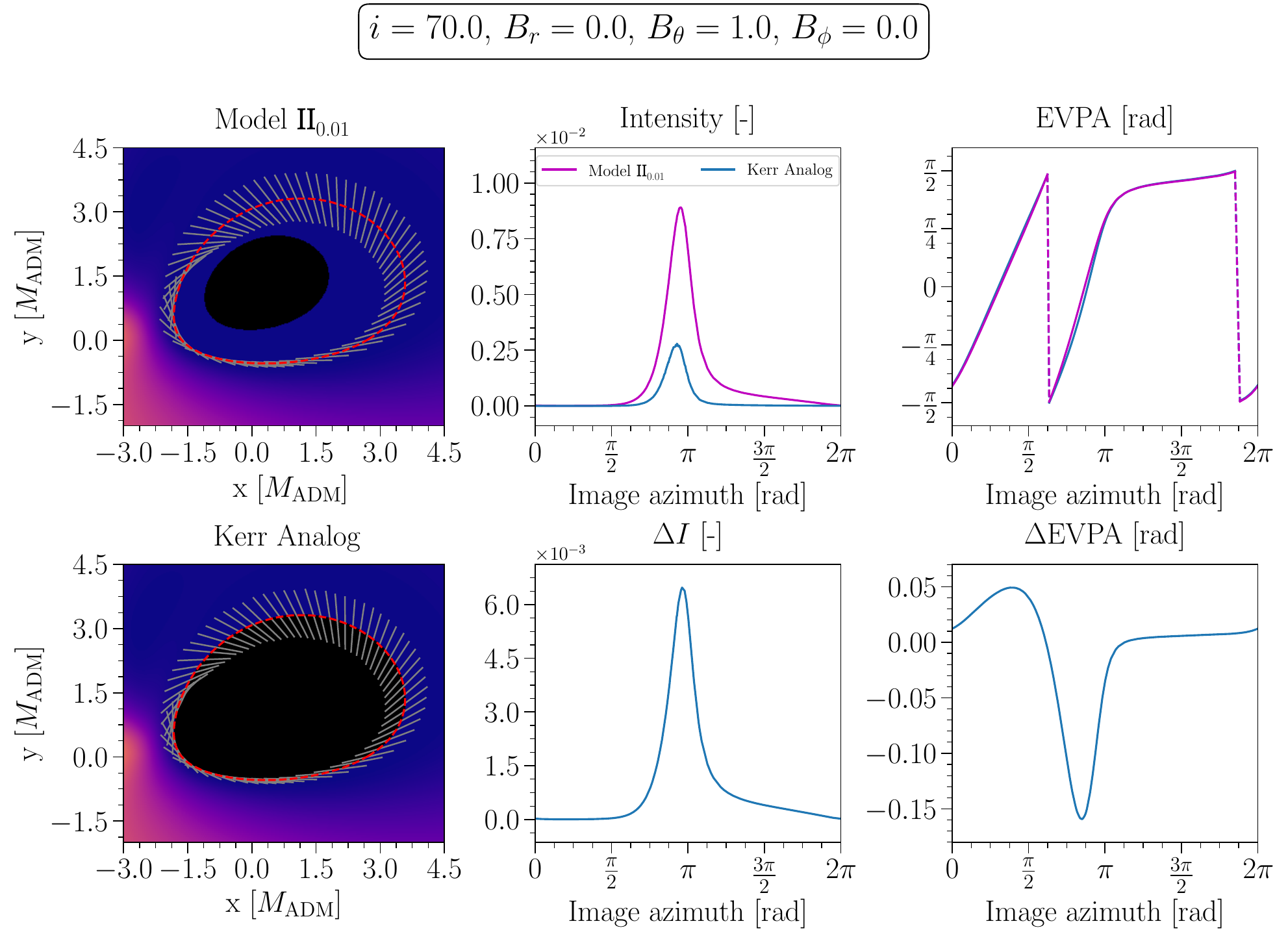}
    \caption{\footnotesize Zoom into the inner regions of the disk from figure \ref{fig:Whole_disk_vert_field_II_V_70_deg}. The leftmost panels show the polarization pattern across the apparent image of the Kerr analog \ac{ISCO}, shown with a red dashed line. The top middle panel shows the intensity variation across this apparent image, going anti-clockwise and starting from the positive $x$-axis. The bottom middle panel shows the difference $\Delta I = I_{\mathbf{II}_{0.01}} - I_{\text{Kerr}}$. The right most panels are analogous, but for the \ac{EVPA}.}
    \label{fig:ISCO_zoom_vert_field_II_70_deg}
\end{figure}

\begin{figure}[H]
    \centering
    \includegraphics[width=1\linewidth]{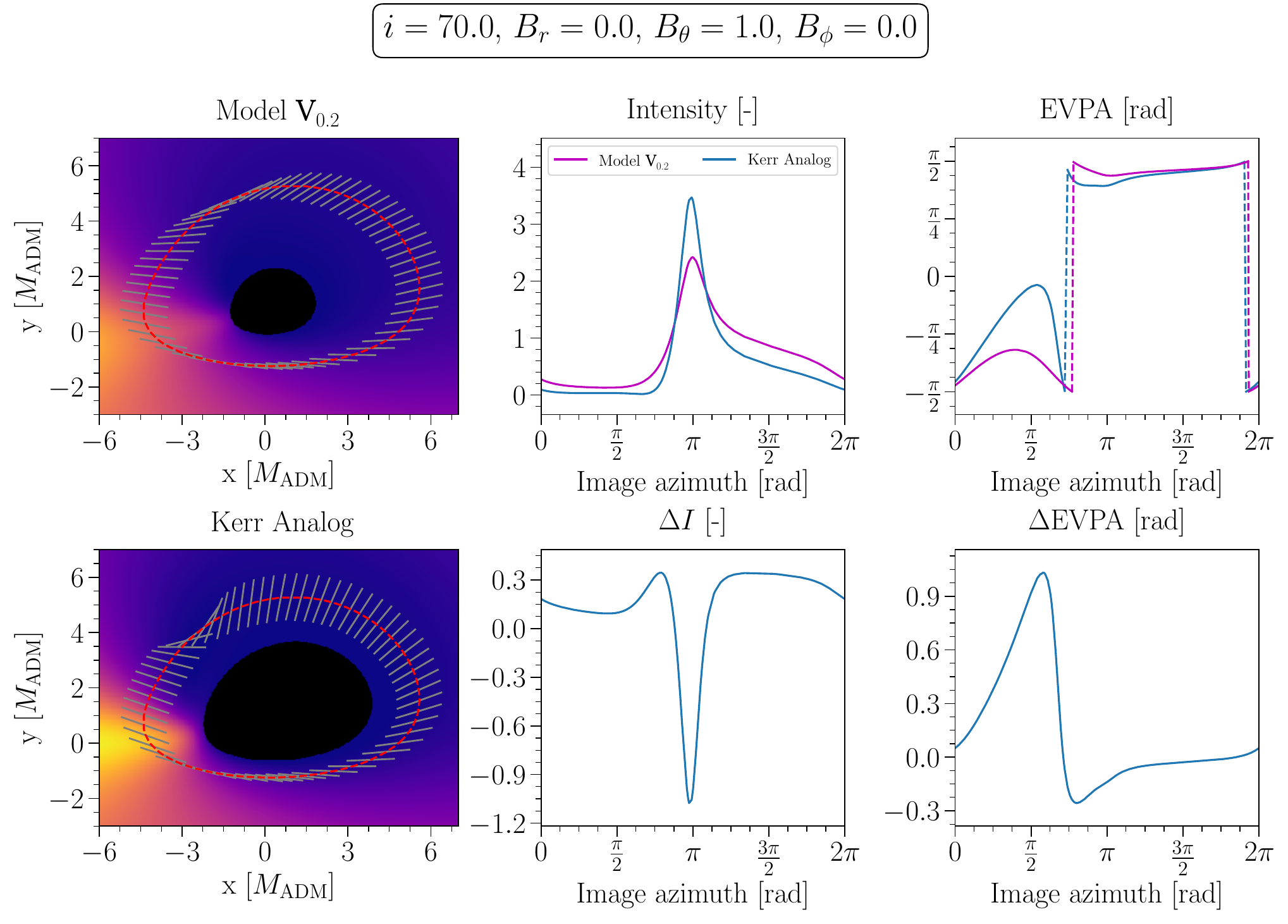}
    \caption{\footnotesize Zoom into the inner regions of the disk from figure \ref{fig:Whole_disk_vert_field_II_V_70_deg}. The leftmost panels show the polarization pattern across the apparent image of the Kerr analog orbit with radius $R = \frac{5}{2}R_\textbf{Kerr}^\text{ISCO}$, shown with a red dashed line. The top middle panel shows the intensity variation across this apparent image, going anti-clockwise and starting from the positive $x$-axis. The bottom middle panel shows the difference $\Delta I = I_{\mathbf{V}_{0.2}} - I_{\text{Kerr}}$. The right most panels are analogous, but for the \ac{EVPA}. }
    \label{fig:2.5_ISCO_zoom_vert_field_V_70_deg}
\end{figure}

\begin{figure}[H]
    \centering
    \includegraphics[width=1\linewidth]{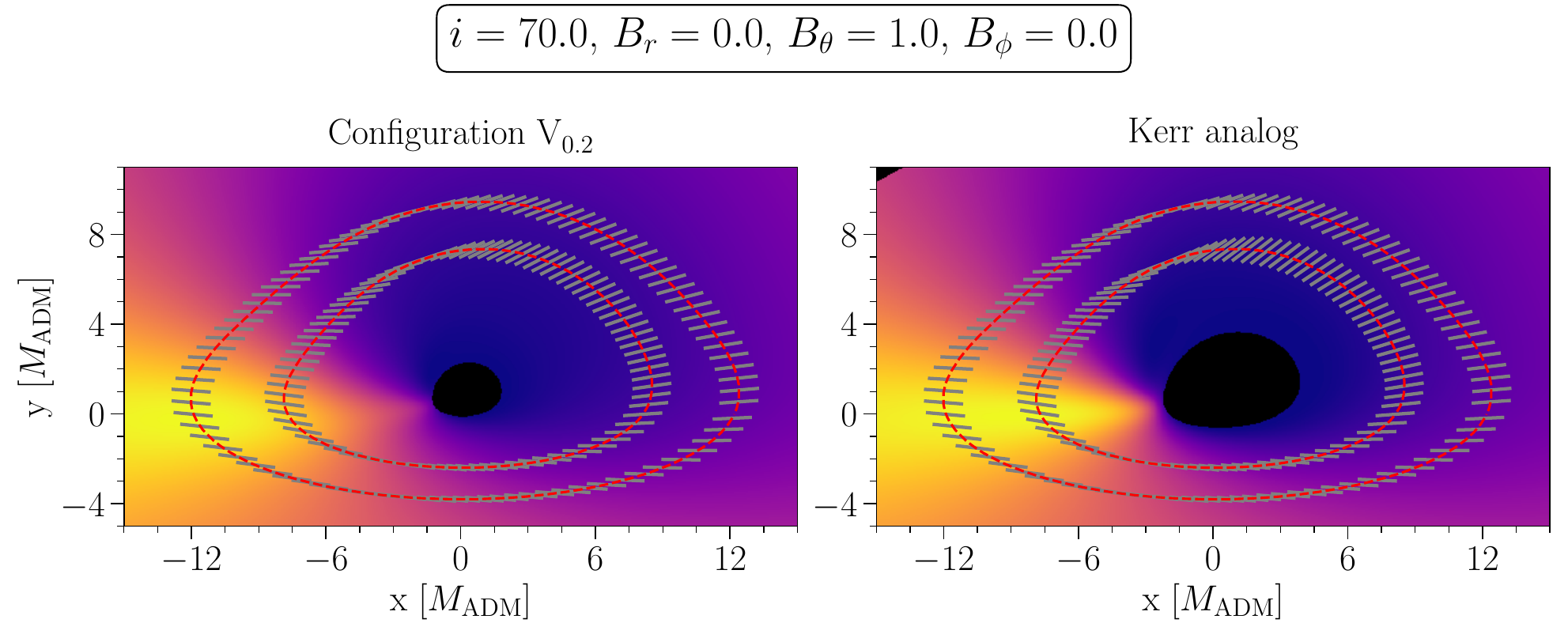}
    \caption{\footnotesize Zoom into the inner regions of the disk from figure \ref{fig:Whole_disk_vert_field_II_V_70_deg}. We plot the polarization pattern across the apparent images of the Kerr analog orbits with radii $R = \{5R^\text{ISCO}_\text{Kerr}, 8R^\text{ISCO}_\text{Kerr}\}$.}
    \label{fig:5_8_ISCO_zoom_vert_field_V_70_deg}
\end{figure}

\begin{figure}[H]
    \centering
    \includegraphics[width=1\linewidth]{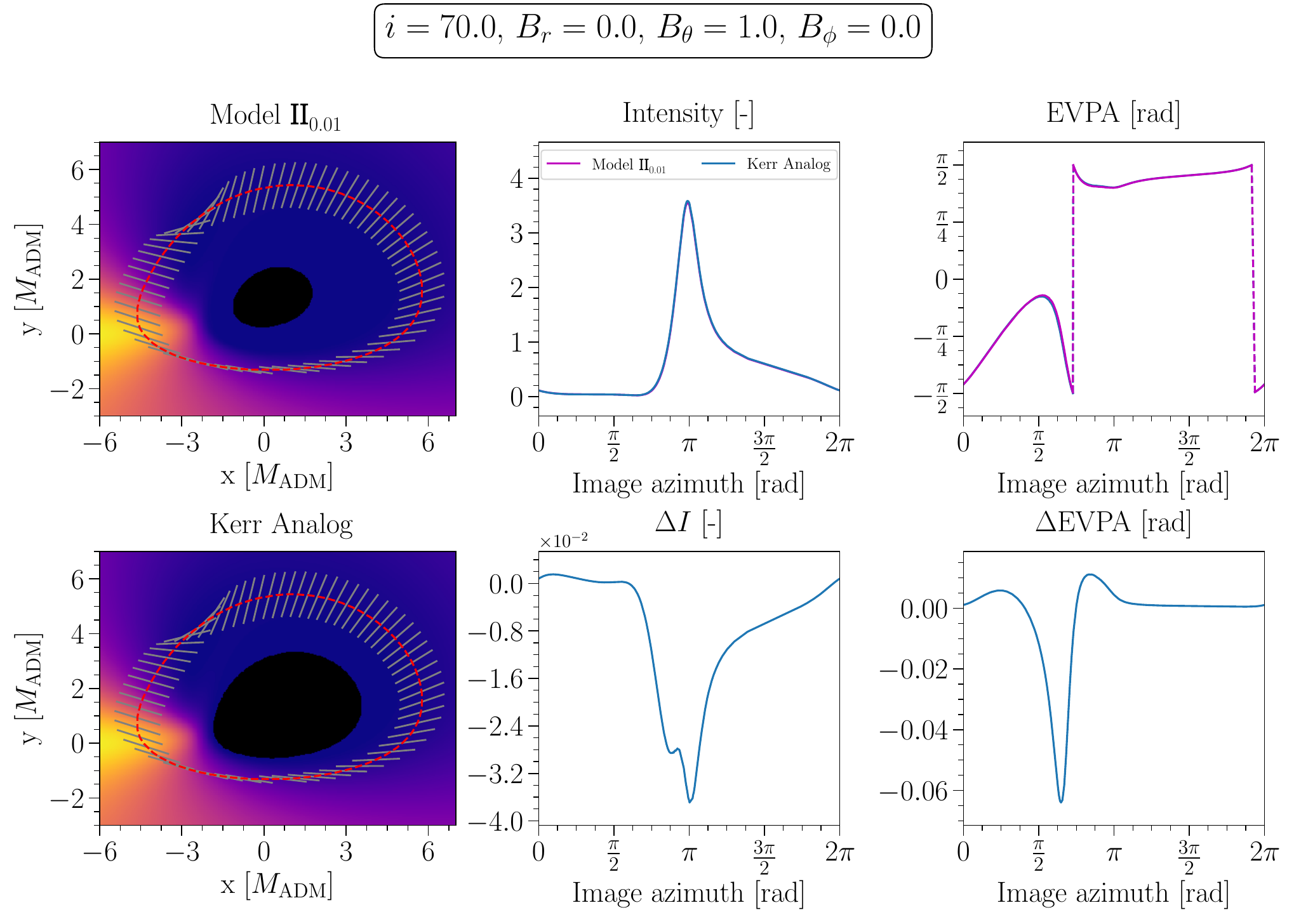}
    \caption{\footnotesize Zoom into the inner regions of the disk from figure \ref{fig:Whole_disk_vert_field_II_V_70_deg}. The leftmost panels show the polarization pattern across the apparent image of the Kerr analog orbit with radius $R = \frac{7}{2}R_\textbf{Kerr}^\text{ISCO}$, shown with a red dashed line. The top middle panel shows the intensity variation across this apparent image, going anti-clockwise and starting from the positive $x$-axis. The bottom middle panel shows the difference $\Delta I = I_{\mathbf{II}_{0.01}} - I_{\text{Kerr}}$. The right most panels are analogous, but for the \ac{EVPA}.}
    \label{fig:3.5_ISCO_zoom_vert_field_II_70_deg}
\end{figure}

Furthermore, if we plot the polarization pattern across the apparent image of the scalarized \ac{ISCO}, rather than across that of its Kerr analog (right panel of figure \ref{fig:ISCO_zoom_vert_field_V_70_deg_2}), the reversal of the \ac{EVPA} disappears. 
The implication of this is that the emission radii play an important role in the effect. We therefore perform this analysis for larger orbital radii in figures \ref{fig:2.5_ISCO_zoom_vert_field_V_70_deg} and \ref{fig:3.5_ISCO_zoom_vert_field_II_70_deg} displaying the polarization pattern for models $\mathbf{II}_{0.01}$ and $\mathbf{V}_{0.2}$ respectively. We choose the Kerr analog orbits' radii such as to best visualize the presence of the effect further from the apparent inner edge of the disk. The strength of this \ac{EVPA} reversal effect diminishes as one goes radially outwards along the image. This can be seen in figure \ref{fig:5_8_ISCO_zoom_vert_field_V_70_deg}, where we plot the polarization pattern for model $\mathbf{V}_{0.2}$ across the apparent position of the Kerr analog orbits with even larger radii $R = \{5R^\text{ISCO}_\text{Kerr}, 8R^\text{ISCO}_\text{Kerr}\}$.\bigskip\par

We find that the reversal of the polarization vector twist direction is also present for the Kerr black holes, but it manifests only at orbital radii, sufficiently far away from the \ac{ISCO}. This also explains its absence in figure \ref{fig:ISCO_zoom_vert_field_V_70_deg_2}. From figures \ref{fig:2.5_ISCO_zoom_vert_field_V_70_deg} and \ref{fig:3.5_ISCO_zoom_vert_field_II_70_deg} we also see that our findings so far also hold for larger orbital radii -- the main imprint of the scalar field on the polarization pattern (which is independent of the intrinsic emissivity of the disk) is a dephasing in the \ac{EVPA}, and this effect is stronger for the less scalarized solutions.

 The fact that this effect is only observed for vertical magnetic fields, and also manifests for Kerr black holes with no scalar field, shows that it is more characteristic of the magnetic field structure than the lensing properties of the underlying spacetime.\bigskip

 We can also note the intensity difference between models \textbf{II}$_{0.01}$ and \textbf{V}$_{0.2}$ on figures \ref{fig:ISCO_zoom_vert_field_V_70_deg} and \ref{fig:ISCO_zoom_vert_field_II_70_deg}. This is similar to the low inclination case, but even stronger -- nearly three orders of magnitude. We again stress that this comparison across models is misleading due to the different spin parameters of the (near extremal) Kerr analogs, which results in large differences in the \ac{ISCO} radii. In fact we showed in \cite{Galin_2026} that at high inclinations, model \textbf{II}$_{0.01}$ produces a significantly higher apparent energy flux from its disk than model \textbf{V}$_{0.2}$ -- around two orders of magnitude at an $80^\circ$ inclination. This is exactly the opposite of what naive comparisons of the intensity values in Figures \ref{fig:ISCO_zoom_vert_field_V_70_deg} and \ref{fig:ISCO_zoom_vert_field_II_70_deg} would suggest. In fact, comparing the intensity values (as computed from \ref{Intensity_definition}) of a numerical model with its own Kerr analog at high inclinations is also misleading, in light of the emission model used in this paper not taking the radial flux distribution into account (as we discussed in section \ref{Subsec:Radial_flux_comment}) and the left panel of figure \ref{fig:ISCO_zoom_vert_field_V_70_deg_2}. For this reason our discussions for the high inclination case focuses on the \ac{EVPA}, rather than the intensity.
 
\newpage

\subsection{Equatorial magnetic fields at low inclination}

\begin{figure}[h]
    \centering
    \includegraphics[width=1\linewidth]{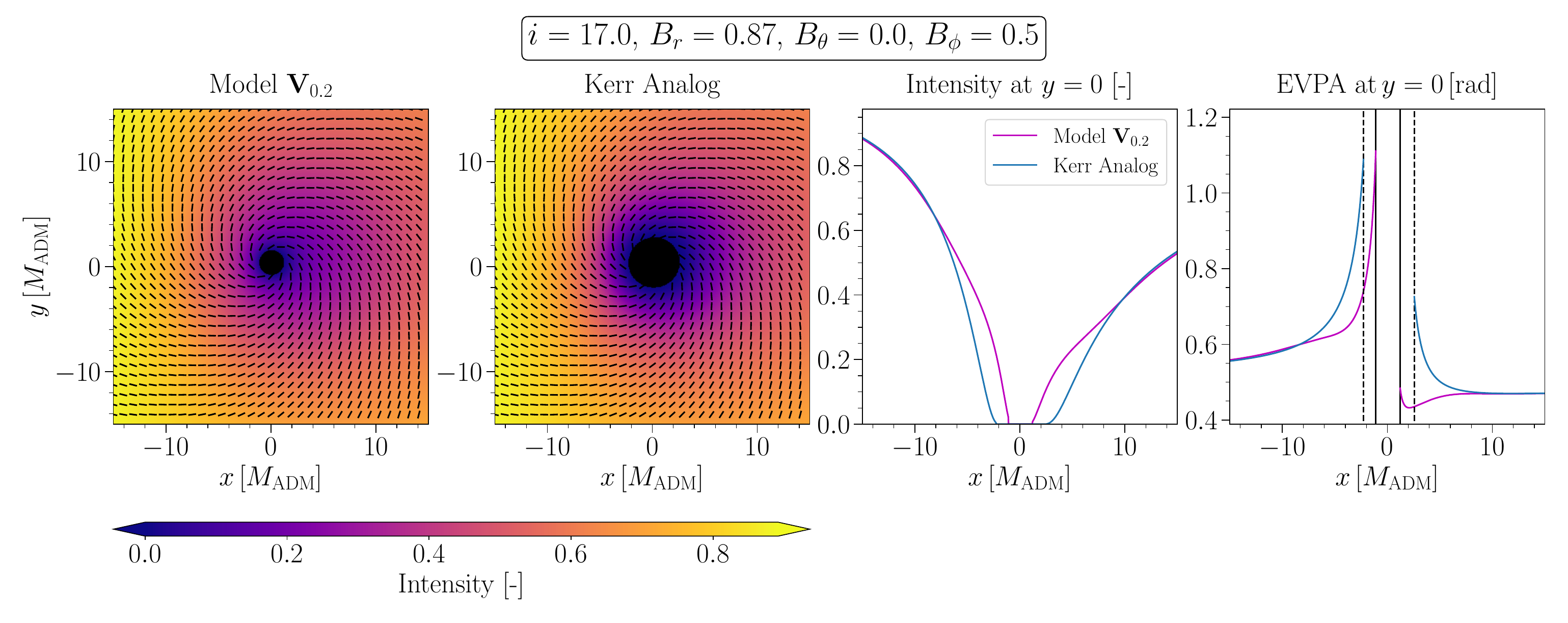}
    \includegraphics[width=1\linewidth]{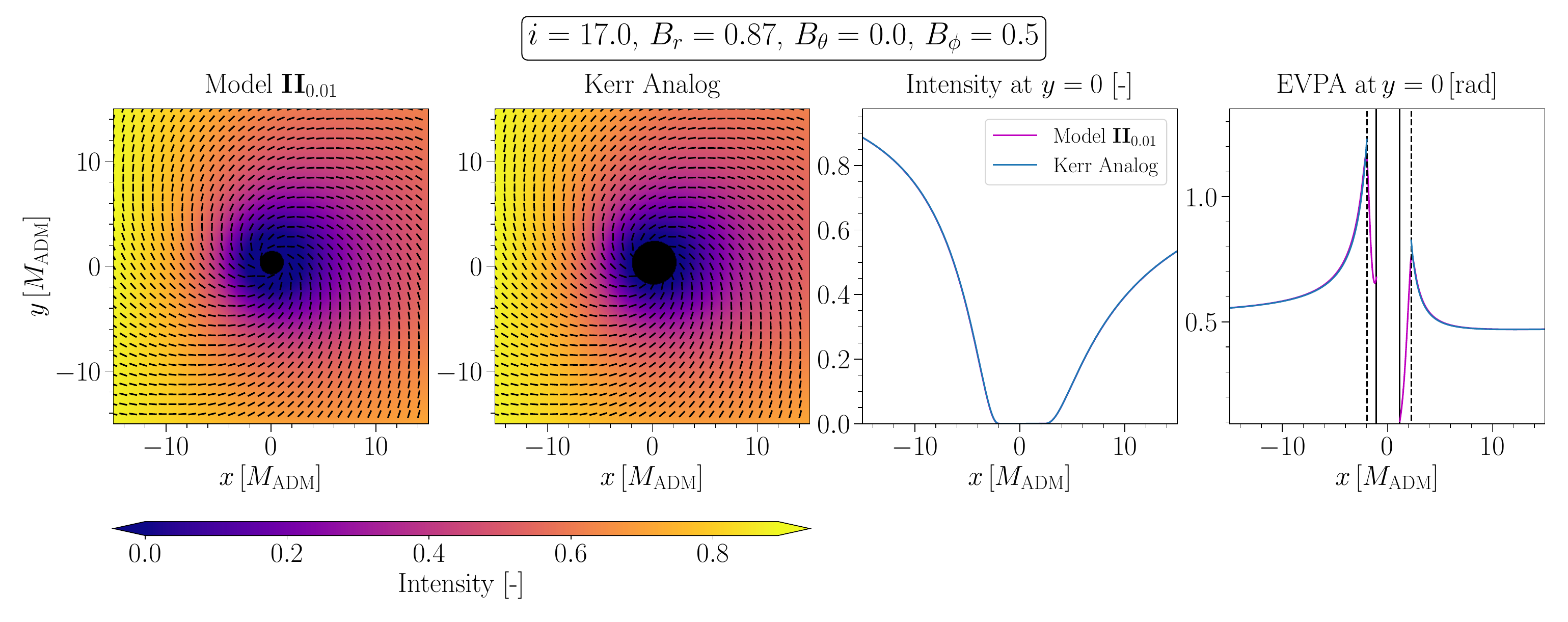}
    \caption{\footnotesize Polarized direct image of a thin accretion disk around the numerical models \textbf{II}$_{0.01}$, \textbf{V}$_{0.2}$ and their respective Kerr analogs for an equatorial magnetic field, viewed at a $17^\circ$ inclination. The two leftmost panels show the overall polarization pattern and its intensity. On the third panel we show the intensity profile as per \eqref{Intensity_definition} at the slice $y = 0$. On the rightmost panel we show the EVPA on the same $y = 0$ slice. The vertical solid/dashed black lines correspond to the apparent inner edge of the numerical/Kerr solutions.}
    \label{fig:Whole_disk_eq_field_II_V_17_deg}
\end{figure}

We now turn our attention to equatorial magnetic fields. Of the three different such field configurations that we have considered, we choose to show $\vec{B} =(0.87, 0, 0.5)$ as a representative example, as we do not see qualitative differences between them. In addition, this field was found to better describe the 2017 polarized observations of M87* \cite{Narayan_2021} (albeit when considering a fluid four-velocity that has a negative radial component, while we consider purely Keplerian motion). Analogously to the previous sections, on figure \ref{fig:Whole_disk_eq_field_II_V_17_deg} we show the overall polarization pattern (two leftmost panels). The third panels show the intensity slices across the $y = 0$ axis of the images and the rightmost panels again show the \ac{EVPA} slices.

We continue to observe that the largest deviations in the intensity slices with respect to the Kerr analogs occur for the numerical models \textbf{V}$_{0.2}$ and \textbf{VI}$_{0.3}$. We can see from the third panel in figure \ref{fig:Whole_disk_eq_field_II_V_17_deg} that the intensity slice for model \textbf{II}$_{0.01}$ is, similarly to the vertical field case, essentially identical to that of its Kerr analog, while the less scalarized model \textbf{V}$_{0.2}$ shows visible deviations. We also notice that the intensity computed from \ref{Intensity_definition} is significantly larger than the vertical field case (both for the scalarized models and their Kerr analogs). This is rather expected, as the intensity scales as $I\propto\sin^2\zeta$, where $\zeta$ is the angle between the photon wave vector and magnetic field as measured by an observer, comoving with the emitting particle. For equatorial fields at low inclinations this angle is naturally closer to $\frac{\pi}{2}$, than for vertical fields.\bigskip\par

From the leftmost panels in figure \ref{fig:Whole_disk_eq_field_II_V_17_deg}, we again see that the overall \ac{EVPA} pattern across the disk remains morphologically similar between the scalarized models and their Kerr analogs. Interestingly, model $\mathbf{II}_{0.01}$ shows a sharp decrease in the \ac{EVPA} around the inner edge of its Kerr analog disk. This is not clearly viable in the large scale pattern, but is noticeable on the rightmost panel.\bigskip\par

\begin{figure}[H]
    \centering
    \includegraphics[width=1\linewidth]{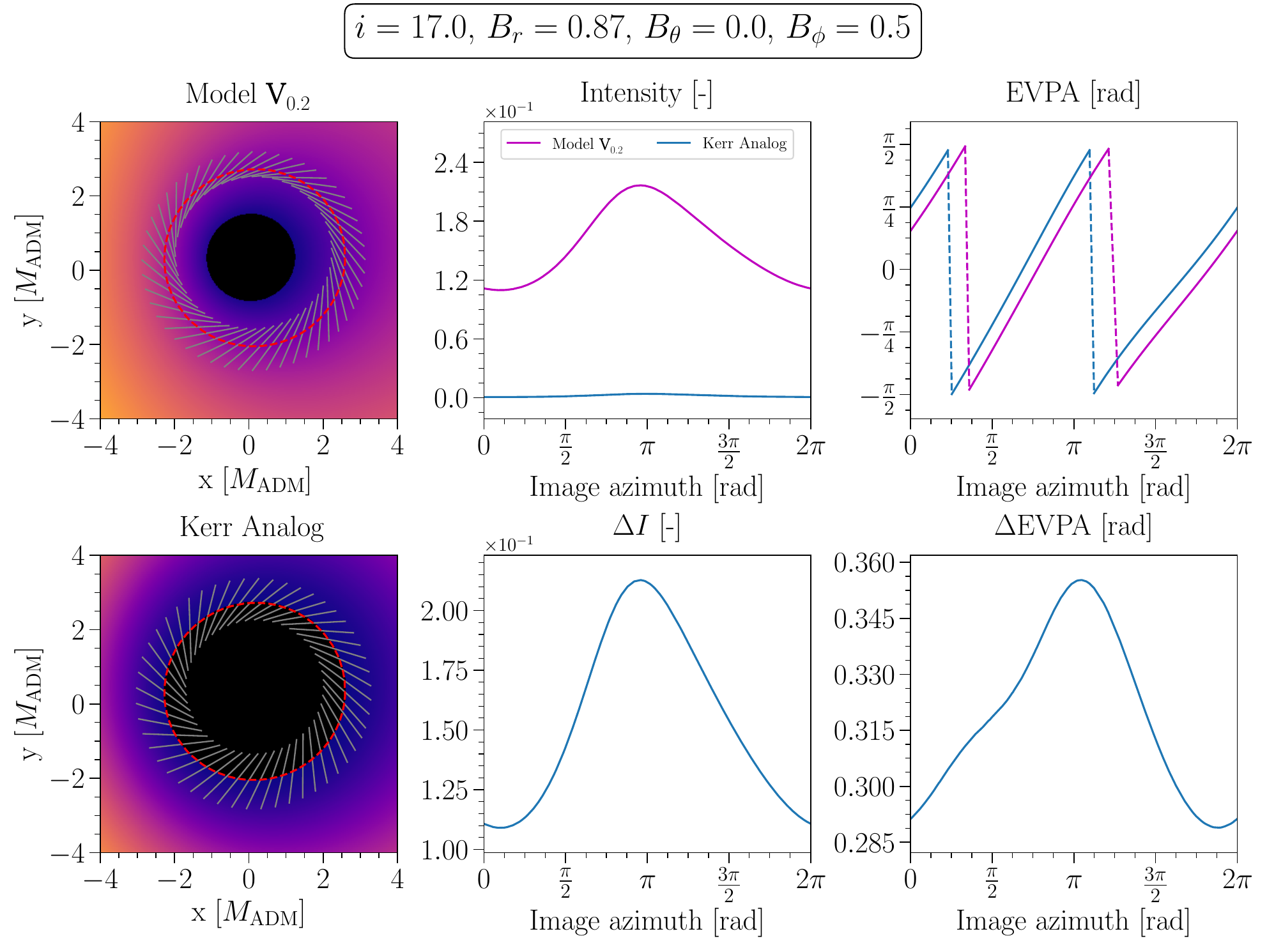}
    \caption{\footnotesize Zoom into the inner regions of the disk from figure \ref{fig:Whole_disk_eq_field_II_V_17_deg}. The leftmost panels show the polarization pattern across the apparent image of the Kerr analog \ac{ISCO}, depicted with a red dashed line. The top middle panel shows the intensity variation across the apparent image, going anti-clockwise and starting from the positive $x$-axis. The bottom middle panel shows the difference $\Delta I = I_{\mathbf{V}_{0.2}} - I_{\text{Kerr}}$. The right most panels are analogous, but for the \ac{EVPA}.}
    \label{fig:ISCO_zoom_eq_field_V_17_deg}
\end{figure}

\begin{figure}[H]
    \centering
    \includegraphics[width=1\linewidth]{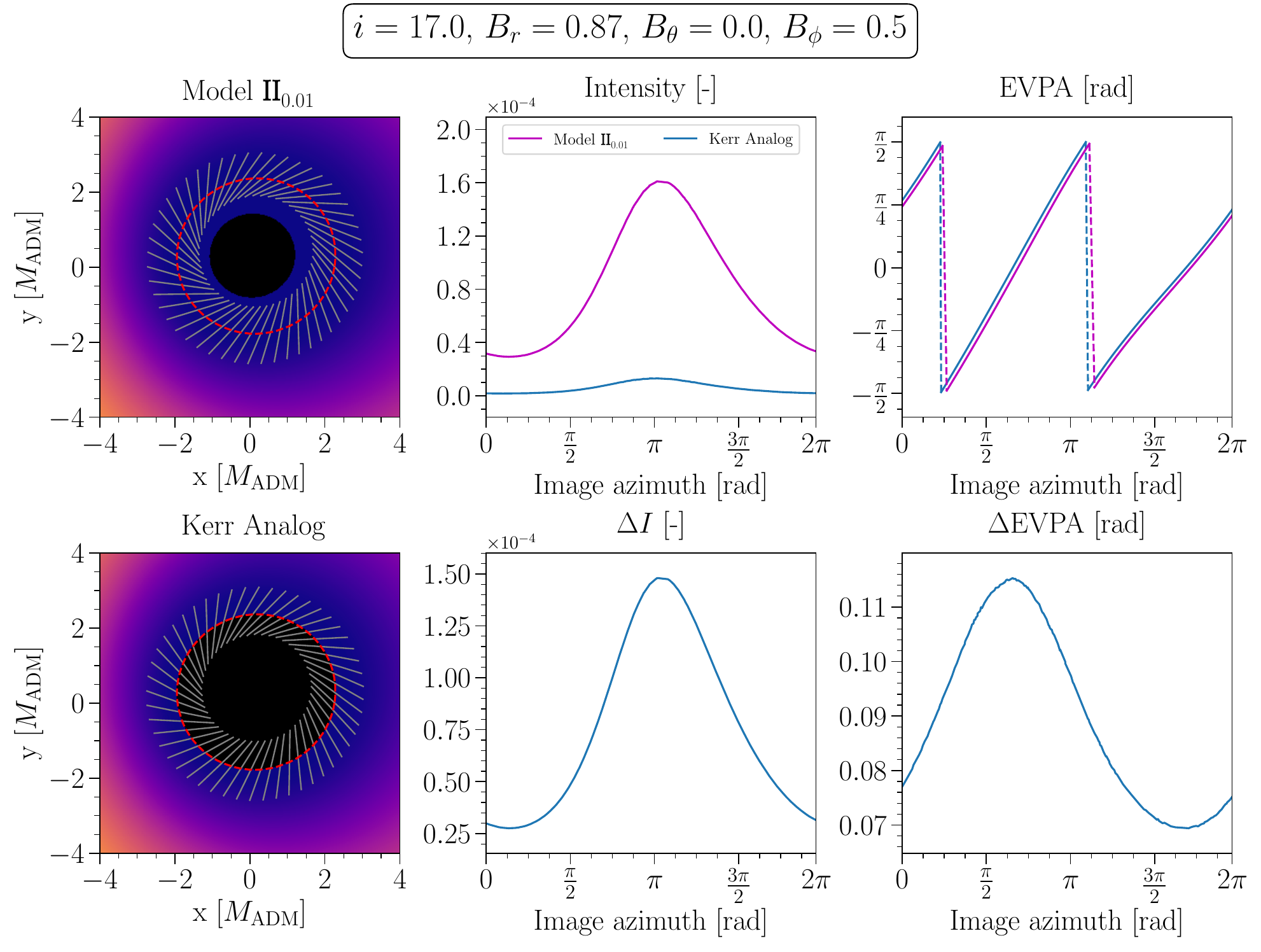}
    \caption{\footnotesize Zoom into the inner regions of the disk from figure \ref{fig:Whole_disk_eq_field_II_V_17_deg}. The leftmost panels show the polarization pattern across the apparent image of the Kerr analog \ac{ISCO}, shown with a red dashed line. The top middle panel shows the intensity variation across this apparent image, going anti-clockwise and starting from the positive $x$-axis. The bottom middle panel shows the difference $\Delta I = I_{\text{II}_{0.01}} - I_{\text{Kerr}}$. The right most panels are analogous, but for the \ac{EVPA}.}
    \label{fig:ISCO_zoom_eq_field_II_17_deg}
\end{figure}

On figures \ref{fig:ISCO_zoom_eq_field_V_17_deg} and \ref{fig:ISCO_zoom_eq_field_II_17_deg} we again show a zoom into the inner regions of the disk. For equatorial fields, model $\mathbf{II}_{0.01}$ shows an even more pronounced decrease in intensity, compared to model $\mathbf{V}_{0.2}$ -- nearly three orders of magnitude. While this is a striking difference, we again stress that the observational consequences of this rely strongly on the radial flux distribution, which we do not take into account. Finally from the rightmost panels we see that, similar to the low inclination vertical field case, the main difference in the \ac{EVPA} across the image of the Kerr analog \ac{ISCO} is dephasing, which is largest for the two least scalarized solutions we consider.

\subsection{Equatorial magnetic fields at high inclination}
\begin{figure}[H]
    \centering
    \includegraphics[width=1\linewidth]{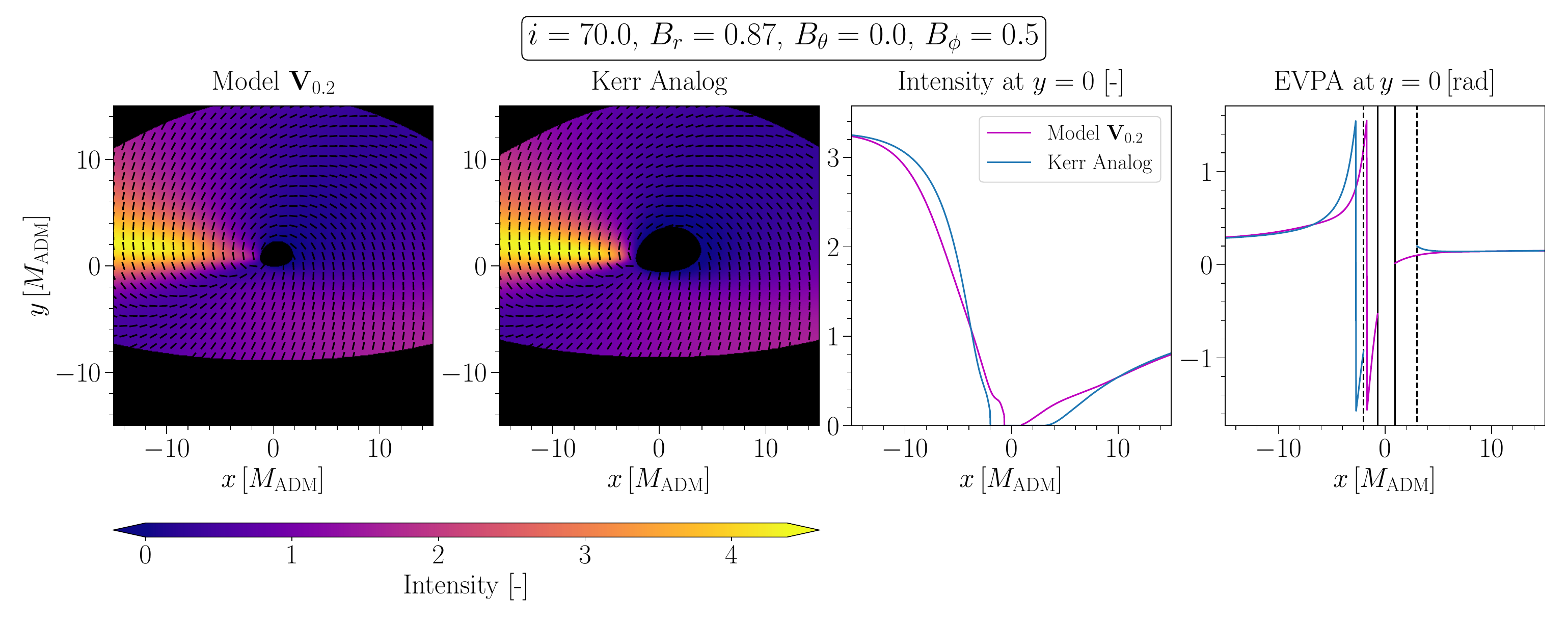}
    \caption{\footnotesize Polarized direct image of a thin accretion disk around the numerical solution $\mathbf{V}_{0.2}$ and its Kerr analog for an equatorial magnetic field, viewed at a $70^\circ$ inclination. On the third panel we show the intensity profile as per \eqref{Intensity_definition} at the slice $y = 0$. On the rightmost panel we show the EVPA on the same $y = 0$ slice. The vertical solid/dashed black lines correspond to the apparent inner edge of the numerical/Kerr solutions.}
    \label{fig:Whole_disk_eq_field_V_70_deg}
\end{figure}

\begin{figure}[H]
    \centering
    \includegraphics[width=0.95\linewidth]{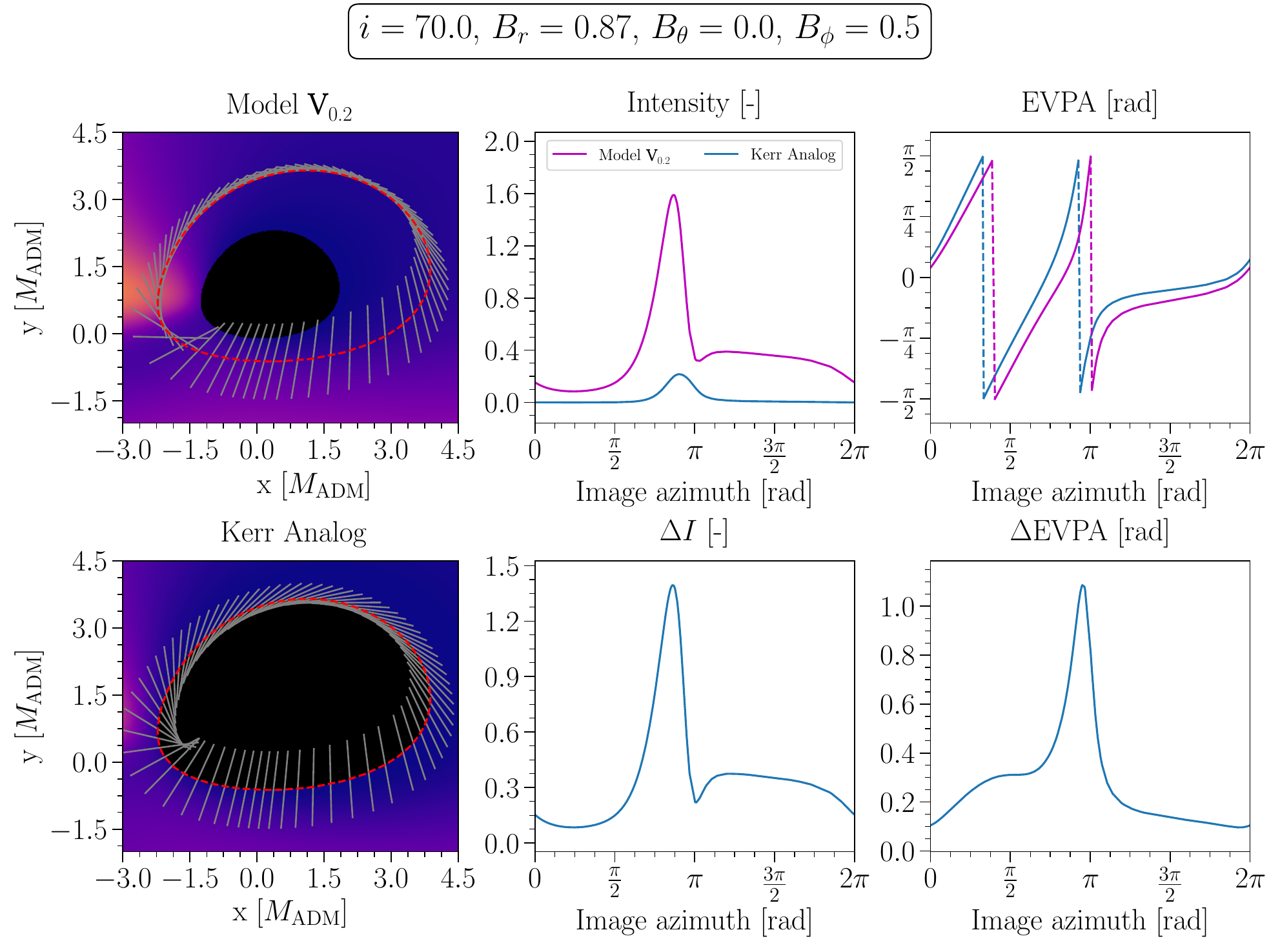}
    \caption{\footnotesize Zoom into the inner regions of the disk from figure \ref{fig:Whole_disk_eq_field_V_70_deg}. The red dashed line corresponds to the apparent shape of the \ac{ISCO} orbit of the Kerr analog. The top middle panel shows the intensity variation across the apparent image, going anti-clockwise and starting from the positive $x$-axis. The bottom middle panel shows the difference $\Delta I = I_{\mathbf{V}_{0.2}} - I_{\text{Kerr}}$. The right most panels are analogous, but for the \ac{EVPA}.}
    \label{fig:ISCO_zoom_eq_field_V_70_deg}
\end{figure}

\begin{figure}[H]
    \centering
    \includegraphics[width=1\linewidth]{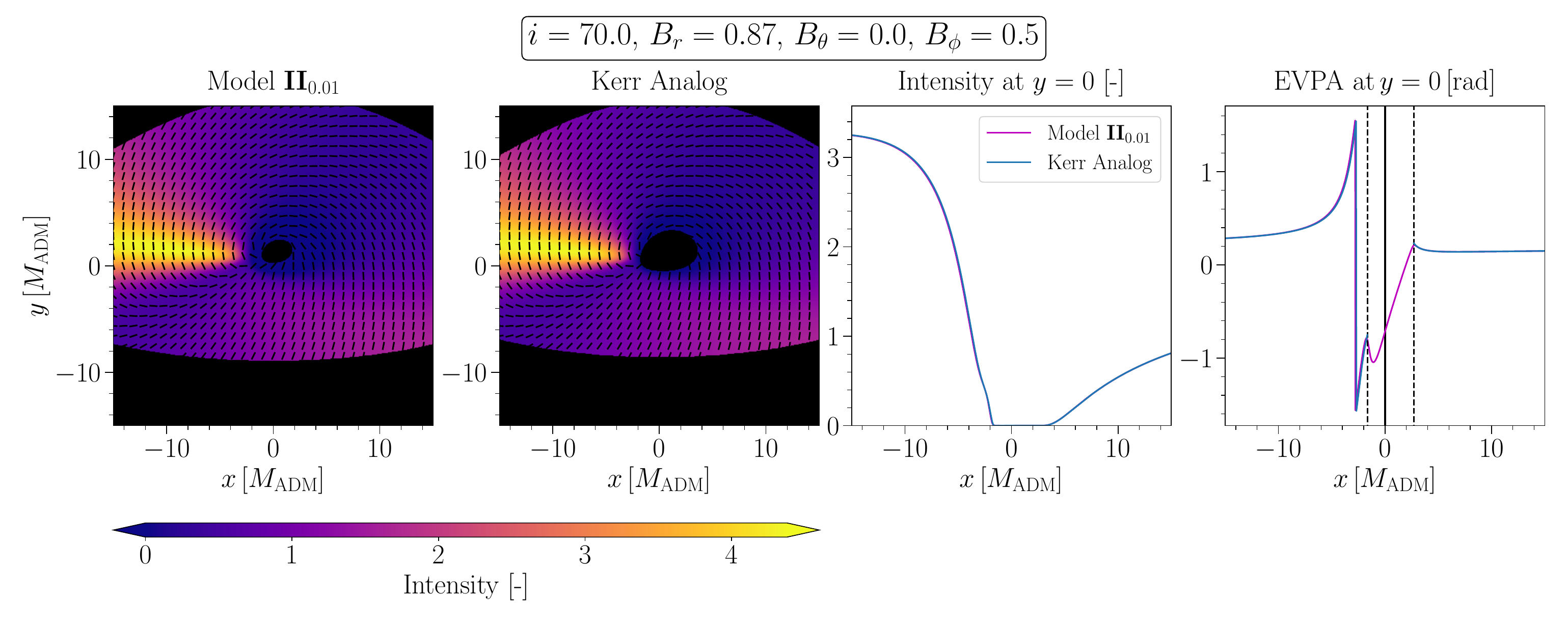}
    \caption{\footnotesize Polarized direct image of a thin accretion disk around the numerical solution \textbf{II}$_{0.01}$ and its Kerr analog for an equatorial magnetic field, viewed at a $70^\circ$ inclination. On the third panel we show the intensity profile as per \eqref{Intensity_definition} at the slice $y = 0$. On the rightmost panel we show the EVPA on the same $y = 0$ slice. The vertical solid/dashed black lines correspond to the apparent inner edge of the numerical/Kerr solutions.}
    \label{fig:Whole_disk_eq_field_II_70_deg}
\end{figure}

\begin{figure}[H]
    \centering
    \includegraphics[width=0.95\linewidth]{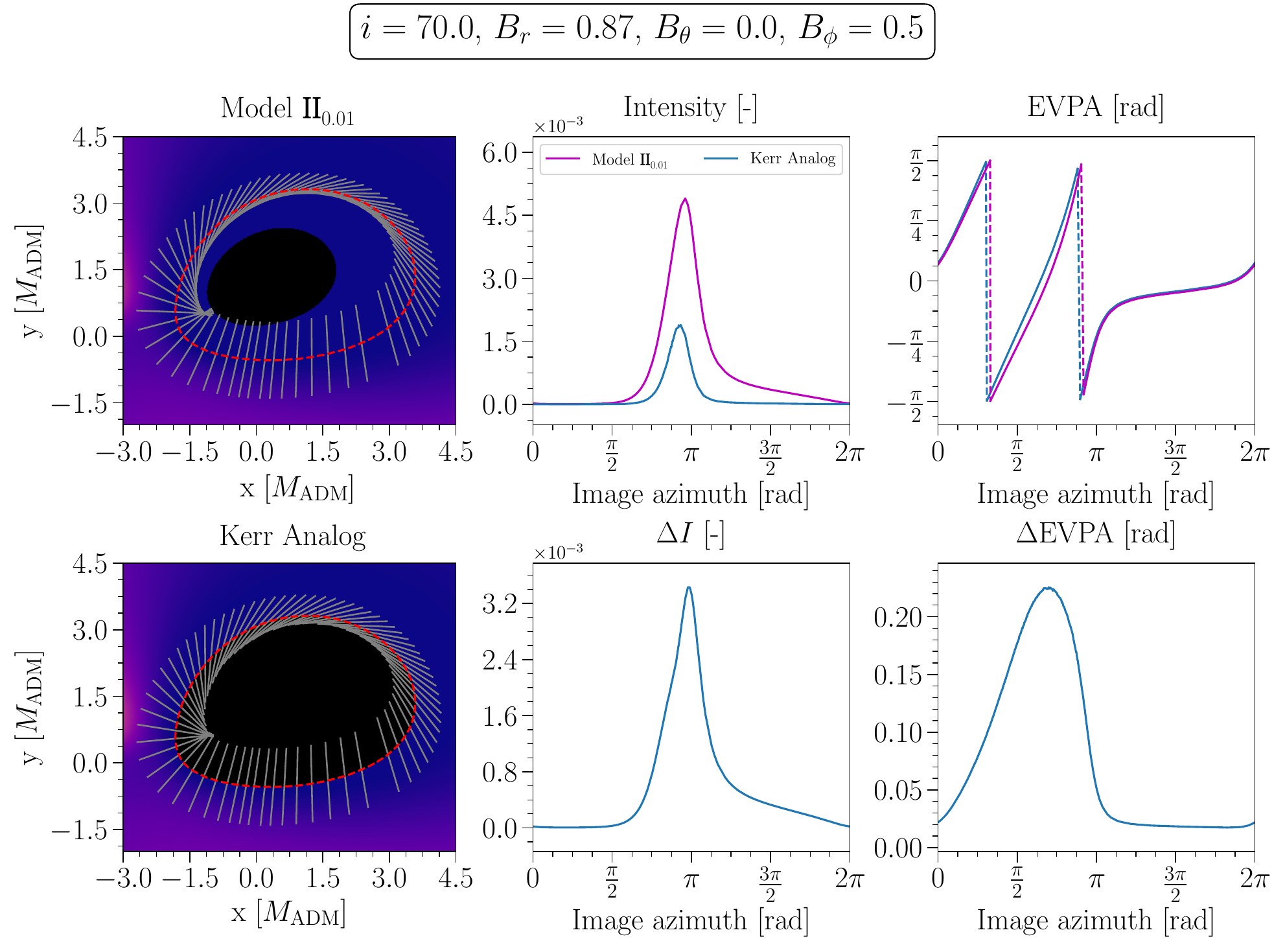}
    \caption{\footnotesize Zoom into the inner regions of the disk from figure \ref{fig:Whole_disk_eq_field_II_70_deg}. The red dashed line corresponds to the apparent shape of the \ac{ISCO} orbit of the Kerr analog. The top middle panel shows the intensity variation across the apparent image, going anti-clockwise and starting from the positive $x$-axis. The bottom middle panel shows the difference $\Delta I = I_{\text{II}_{0.01}} - I_{\text{Kerr}}$. The right most panels are analogous, but for the \ac{EVPA}.}
    \label{fig:ISCO_zoom_eq_field_II_70_deg}
\end{figure}

We now move on to the high inclination $i = 70^\circ$ case for equatorial magnetic fields. Analogously to the above sections, in figures \ref{fig:Whole_disk_eq_field_V_70_deg} and \ref{fig:Whole_disk_eq_field_II_70_deg} we show the images of the entire disk, with the corresponding intensity and \ac{EVPA} slices, while on figures \ref{fig:ISCO_zoom_eq_field_V_70_deg} and \ref{fig:ISCO_zoom_eq_field_II_70_deg} we show a zoom of the inner region and the polarization pattern across the apparent position of the Kerr analog \ac{ISCO}.\bigskip\par

From figures \ref{fig:Whole_disk_eq_field_V_70_deg} and \ref{fig:Whole_disk_eq_field_II_70_deg} we again find no large scale morphological differences in the polarization pattern across the disk. The relationship of the highest deviations in the intensity and \ac{EVPA} slices corresponding to the least scalarized solutions $\mathbf{V}_{0.2}$ and $\mathbf{VI}_{0.3}$ continues to hold. From figures \ref{fig:ISCO_zoom_eq_field_V_70_deg} and \ref{fig:ISCO_zoom_eq_field_II_70_deg} we notice, unlike the vertical field case, a lack of the distinct reversal of the \ac{EVPA}. This shows that the effect is largely magnetic field dependent.

\newpage

\newpage 

\subsection{Physical origin of the EVPA dephasing}\label{EVPA dephasing origin}

In order to determine what causes the observed \ac{EVPA} dephasing, is it helpful to follow the evolution of the polarization vector along particular geodesics. We choose the 70$^\circ$ inclination scenario with a vertical magnetic field $\vec{B} = (0, 1.0, 0)$ as a case study, as there we see the strongest dephasing. The particular geodesics we will follow correspond to the $\Delta\text{EVPA}$ peaks in figures \ref{fig:ISCO_zoom_vert_field_V_70_deg} and \ref{fig:ISCO_zoom_vert_field_II_70_deg}. The less scalarized model $\mathbf{V}_{0.2}$ shows two peaks, which have the following coordinates on the observer's screen: $\{x, y\} = \{ -2.147\,M_\text{ADM},\,1.174\, M_\text{ADM}\}$ and $\{x, y\} = \{-0.282\, M_\text{ADM},\, 3.345\, M_\text{ADM}\}$. The more scalarized model $\mathbf{II}_{0.01}$ also shows two peaks -- a bigger one at coordinates $\{x, y\}=\{-1.792\, M_\text{ADM},\, 0.894\, M_\text{ADM}\}$ and another, smaller one, which we will not consider here. \bigskip\par

\begin{figure}[H]
    \centering
    \begin{subfigure}{\textwidth}
        \centering
        \includegraphics[width=0.8\linewidth]{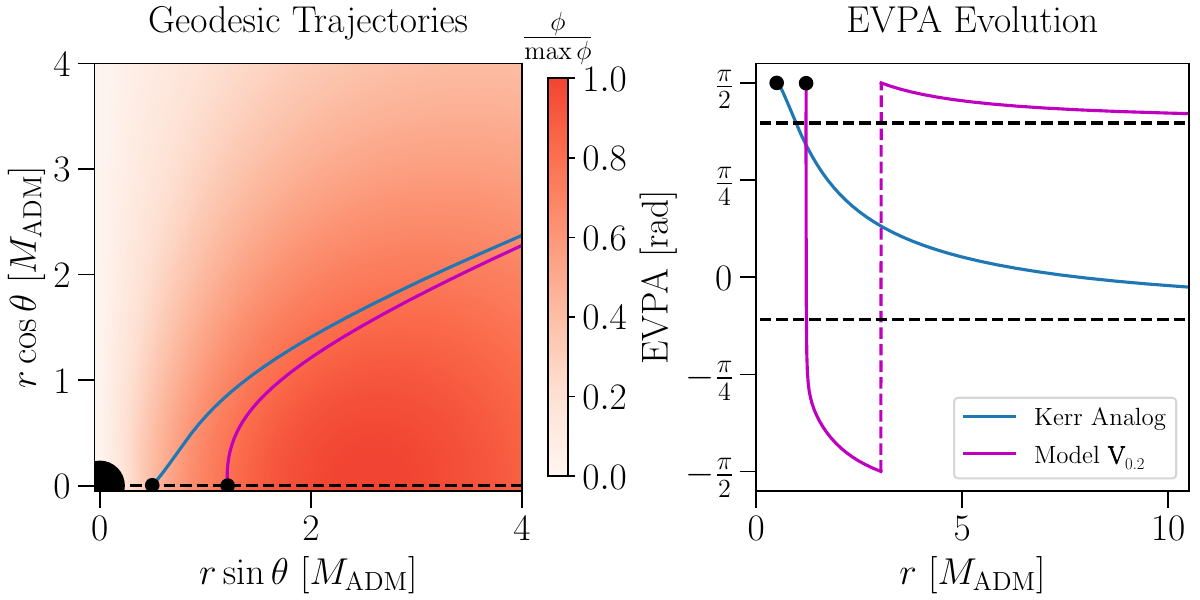}
        \caption{\footnotesize Geodesic with coordinates on the observer's screen $\{x, y \}=\{-2.147\, M_\text{ADM},\,1.174\, M_\text{ADM}\}$} 
    \end{subfigure}
    \begin{subfigure}{\textwidth}
        \centering
        \includegraphics[width=0.8\linewidth]{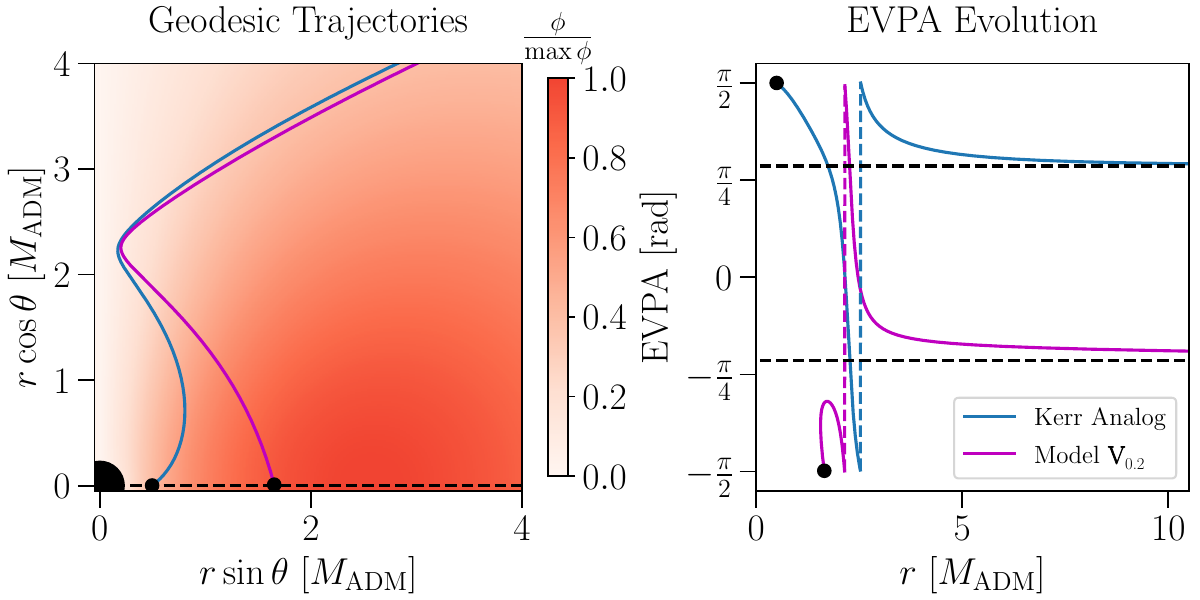}
        \caption{\footnotesize Geodesic with coordinates on the observer's screen $\{x, y \}=\{-0.282\, M_\text{ADM},\,3.345\, M_\text{ADM}\}$}
    \end{subfigure}
    \caption{\footnotesize Plotted trajectory, and \ac{EVPA} evolution along it, for chosen geodesics in the spacetime of model $\mathbf{V}_{0.2}$ and its Kerr analog. The left panel shows the projected trajectory of the photon with the scalar field density overlaid with a colormap. The right panel shows the evolution of the \ac{EVPA} as a function of the radial coordinate. The dotted horizontal lines on the right panels show the asymptotic value of the \ac{EVPA}, while the black dots show the emission point. All trajectories are plotted in the coordinates of line element \eqref{lucas line element}.}
    \label{fig:EVPA_evolution_V_vert_field_p1_p2}
\end{figure}

\begin{figure}[H]
    \centering
    \includegraphics[width=0.8\linewidth]{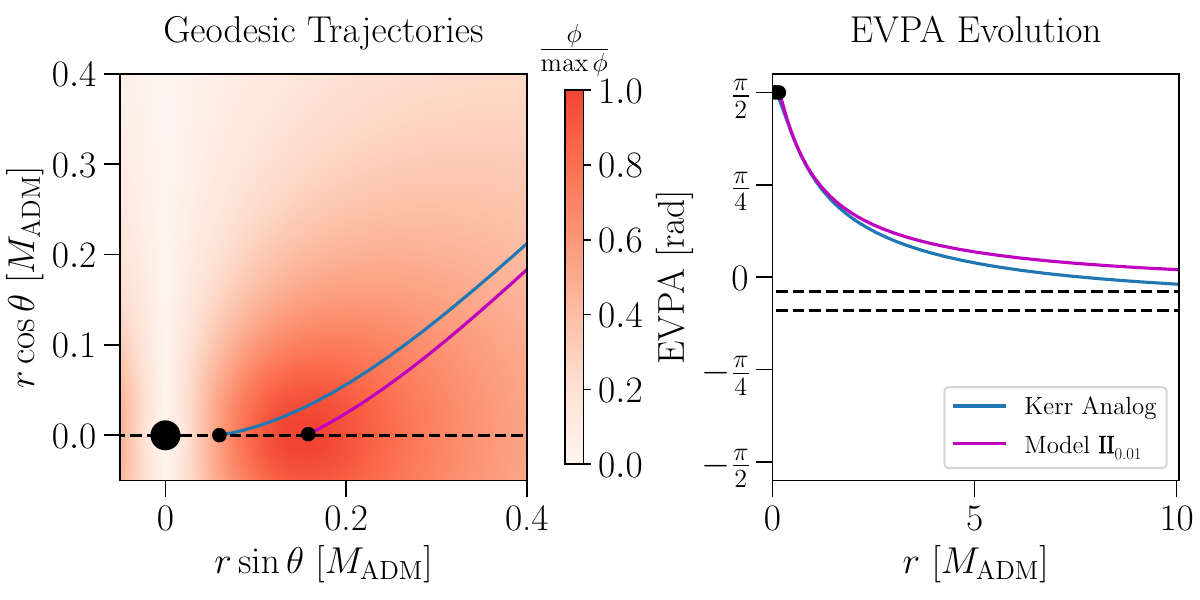}
    \caption{\footnotesize Plotted trajectory, and \ac{EVPA} evolution along it, for chosen geodesics in the spacetime of model $\mathbf{II}_{0.01}$ and its Kerr analog. The left panel shows the projected trajectory of the photon with the scalar field density overlaid with a colormap. The right panel shows the evolution of the \ac{EVPA} as a function of the radial coordinate. The dotted horizontal lines on the right panels show the asymptotic value of the \ac{EVPA}, while the black dots show the emission point. All trajectories are plotted in the coordinates of line element \eqref{lucas line element}. The coordinates of these geodesics on the observer's screen are $\{x, y \}=\{-1.792\, M_\text{ADM},\, 0.894\, M_\text{ADM}\}$. They correspond to the maximum $\Delta \text{EVPA}$ in figure \ref{fig:ISCO_zoom_eq_field_II_70_deg}.}
    \label{fig:EVPA_evolution_II_vert_field_p1}
\end{figure}

The corresponding trajectories and \ac{EVPA} evolution are displayed on figures \ref{fig:EVPA_evolution_V_vert_field_p1_p2} and \ref{fig:EVPA_evolution_II_vert_field_p1}. The left panels show the projection of the trajectories in the ($r\sin\theta  - z$) plane, and the scalar field overlaid with a colormap. The right panel shows the \ac{EVPA} evolution as a function of the radial coordinate of line element \eqref{lucas line element}, as measured by a \ac{ZAMO} at the given point. The black dots correspond to the emission points, which in the chosen examples are the \ac{ISCO}s, while the horizontal dotted black lines are the asymptotic value of the \ac{EVPA}.\bigskip\par

For each chosen geodesic we also follow its corresponding Kerr analog, which would arrive at the same coordinates on the observer's screen. These analog geodesics are very close to their numerical counterparts far away from the black hole, but we can see from the left panels of figure \ref{fig:EVPA_evolution_V_vert_field_p1_p2} that their initial conditions are substantially different for the less scalarized model $\mathbf{V}_{0.2}$. To interpret this physically it is best if one considers backtracing the geodesics from their endpoint (the observer) to their emission point (the \ac{ISCO}). We first notice (again from the left panel of figure \ref{fig:EVPA_evolution_V_vert_field_p1_p2}) that model \textbf{V}$_{0.2}$ has its \ac{ISCO} positioned between the central black hole and the scalar field torus.  The geodesics we choose to follow are emitted from the far side of the black hole, relative to the observer. This means that to reach the observer, they must pass between the black hole and the scalar field torus. As one follows these geodesics back towards their origin, the ones of the numerical model will be pulled towards the scalar field torus, causing them to diverge from their Kerr analog. Given that the process of parallel transport depends strongly on the geodesic path, we can attribute the observed dephasing in the \ac{EVPA} to the different trajectories that the photons take near the \ac{ISCO} in order to reach the observer. \bigskip\par

Given this explanation, one would expect that the front part of the disk would not exhibit such large \ac{EVPA} deviations, owing to the fact that the corresponding geodesics pass through the scalar field, rather than between it and the central black hole. Indeed, if we refer back to figures \ref{fig:ISCO_zoom_vert_field_V_70_deg}, \ref{fig:ISCO_zoom_vert_field_II_70_deg}, \ref{fig:2.5_ISCO_zoom_vert_field_V_70_deg}, \ref{fig:3.5_ISCO_zoom_vert_field_II_70_deg}, \ref{fig:ISCO_zoom_eq_field_V_70_deg}, \ref{fig:ISCO_zoom_eq_field_II_70_deg} we see that the front part of the disk (corresponding to image azimuth values approximately in the range $[\pi, 2\pi]$) shows a plateau near zero on the $\Delta\text{EVPA}$ panel. This clearly demonstrates that the large \ac{EVPA} deviations are caused by the diverging of the geodesics from their Kerr analogs due to the presence of the scalar field cloud. Furthermore, the right panels of figure \ref{fig:EVPA_evolution_V_vert_field_p1_p2} show that this effect can manifest strongly even when the initial polarization vectors start with the same \ac{EVPA}.\bigskip\par

Repeating this analysis for the more strongly scalarized model $\mathbf{II}_{0.01}$ we first find, from the left panel of figure \ref{fig:EVPA_evolution_II_vert_field_p1}, that its \ac{ISCO} lies very close to the scalar field maximum inside the torus. This in fact is the cause of the smaller observed $\Delta\text{EVPA}$. For a (direct) geodesic emitted at the \ac{ISCO} to reach the observer, it never has to pass between the scalar field torus and the black hole. This results in the geodesic deviating minimally from its Kerr analog, which in turn results in a smaller $\Delta\text{EVPA}$.\bigskip\par

We conclude that the unexpected result, showing a higher $\Delta\text{EVPA}$ for solutions closer to Kerr, is due to the relative positioning of the \ac{ISCO} and the scalar field torus. Models with a smaller normalized Noether charge $q$ have their \ac{ISCO} between the torus and the black hole, resulting in a substantially different geodesic path near the emission point. This results in a large observed $\Delta\text{EVPA}$. In contrast to this, models with a higher $q$ have their \ac{ISCO} inside the torus. This results in the direct geodesics not deviating from their Kerr analogs substantially, which in turn results in a smaller $\Delta\text{EVPA}$.

\subsection{Maximum intensity deviation for vertical magnetic fields}\label{Subsec::deviation_vert_intensity}
In table \ref{tab:Deltas_vert_field} we present the maximum intensity deviation $\Delta I$ along the apparent position of two Kerr analog orbits. The orbit radii $R$ are chosen to be $R = \{R^\text{ISCO}_\text{Kerr},\frac{3}{2}R^\text{ISCO}_\text{Kerr}\}$. We choose these two orbits because as seen from table \ref{tab:Kerr_analog_spins}, most Kerr analogs are near extremal. This means that their \ac{ISCO} will be very close to the infinite redshift surface, resulting in a small value for the Doppler factor $g$ appearing in eq. \ref{Intensity_definition}. As commented above, the numerical solutions we consider tend to have a weaker focusing effect on the null geodesics, resulting in larger source radii for given impact parameters (see figure \ref{fig:ISCO_zoom_vert_field_V_70_deg_2}), where the Doppler factor $g$ is larger for a fixed inclination. It is then interesting to look at how the intensity deviations change as we go radially out along the image. We therefore choose to also analyze the apparent position of the $R = \frac{3}{2}R^\text{ISCO}$ Kerr analog orbit.

\begin{table}[htbp]

    \footnotesize
    \setlength{\tabcolsep}{8.7pt}
    \renewcommand{\arraystretch}{1.7}

    \begin{tabular}{ccccccc}
        \hline\hline
        \multicolumn{7}{c}{$B_r = 0.0$ $B_\theta = 1.0$ $B_\phi = 0.0$}\\
        \hline\hline
        &\multicolumn{3}{c}{Apparent image of the orbit with $R = R^\text{ISCO}_\text{Kerr}$} & \multicolumn{3}{c}{Apparent image of the orbit with $R = \frac{3}{2}R^\text{ISCO}_\text{Kerr}$}\\

        \hline\hline
        
        Model & $i_\text{obs} = 17^\circ$ & $i_\text{obs} = 45^\circ$ &  $i_\text{obs} = 70^\circ$ & $i_\text{obs} = 17^\circ$ & $i_\text{obs} = 45^\circ$ &  $i_\text{obs} = 70^\circ$\\ [1.5ex]
        
        \hline
        
        $\mathbf{I}_{\,0.01}$ & - & - & - & $\left(-0.41,1.01\pi\right)$ & $\left(-0.43,0.92\pi\right)$ & $\left(-0.59,0.95\pi\right)$ \\
        $\mathbf{II}_{\,0.01}$ & $\left(10.7,1.05\pi\right)$ & $\left(6.26,0.98\pi\right)$  & $\left(2.76,0.96\pi\right)$ & $\left(-0.12,0.98\pi\right)$ & $\left(-0.18,0.97\pi\right)$ & $\left(-0.23,0.95\pi\right)$ \\
        $\mathbf{III}_{\,0.05}$ & $\left(26.2,1.13\pi\right)$ & $\left(12.2,1.05\pi\right)$ & $\left(5.27,1.00\pi\right)$ & $\left(-0.22,1.00\pi\right)$ & $\left(-0.41,0.97\pi\right)$ & $\left(-0.58,0.95\pi\right)$ \\
        $\mathbf{IV}_{\,0.1}$ & $\left(34.2,1.21\pi\right)$ & $\left(15.9,1.08\pi\right)$ & $\left(6.34,1.01\pi\right)$ & $\left(0.58,1.43\pi\right)$ & $\left(-0.36,0.93\pi\right)$ & $\left(-0.54,0.95\pi\right)$ \\
        $\mathbf{V}_{\,0.2}$ & $\left(30.1,1.42\pi\right)$ & $\left(14.4,1.15\pi\right)$ & $\left(4.39,0.99\pi\right)$ & $\left(2.79,1.64\pi\right)$ & $\left(2.12,1.33\pi\right)$ & $\left(1.39,1.11\pi\right)$ \\
        $\mathbf{VI}_{\,0.3}$ & $\left(1.24,1.39\pi\right)$  & $\left(0.90,1.17\pi\right)$  & $\left(0.52,1.02\pi\right)$  & $\left(0.53,1.76\pi\right)$ & $\left(0.45,1.39\pi\right)$  & $\left(0.49,0.87\pi\right)$  \\
        
        \hline		
        
    \end{tabular}
    \caption{\label{tab:Deltas_vert_field}\footnotesize The maximum intensity deviation $\Delta I$ along the apparent position of two Kerr analog orbits (with radii $R = \{R^\text{ISCO}_\text{Kerr}, \frac{3}{2}R^\text{ISCO}_\text{Kerr}\}$) for all six solutions considered in this paper and for the equatorial magnetic field with components $B_r = 0.0$, $B_\theta = 1.0$ and $B_\phi = 0.0$. The reported intensity values are normalized to the corresponding value of the Kerr analog and are presented as $(\frac{\max \Delta I}{I_\text{Kerr}},\,\phi)$, where $\phi$ is the azimuthal angle at which the maxima occur. For the specific case R = $R^\text{ISCO}_\text{Kerr}$, the apparent position of this orbit does not entirely overlap with the image of the disk for model $\mathbf{I}_{0.01}$. We therefore do not specify maximum intensity deviations for this case.}
    
\end{table}

We see that the intensity deviations are significantly smaller for solution $\mathbf{VI}_{0.3}$ over the apparent image of the Kerr analog \ac{ISCO}. This is largely due to the fact that solution $\mathbf{VI}_{0.3}$'s Kerr analog has a smaller spin parameter than the rest (see table \ref{tab:Kerr_analog_spins}), and thus its \ac{ISCO} is further away from the infinite redshift surface, which leads to a larger intensity (due to the higher numerical value of the Doppler factor $g$ in eq. \ref{Intensity_definition}).\bigskip\par

We also see that for the $R = \frac{3}{2}R^\text{ISCO}$ Kerr analog orbit, the intensity deviations decrease for all numerical solutions we consider and even reverse sign for some of them. Here it is worth stressing that these relative intensity deviations in practice also depends on the intrinsic emissivity of the disk (as discussed in section \ref{Subsec:Radial_flux_comment}). It is expected that this emissivity is a function that varies rapidly with $r_\text{source}$ for radii close to the \ac{ISCO} \cite{Collodel_2021}. Given the relatively large spread of $r_\text{source}$ over the apparent image of a Kerr analog orbit at high inclinations (see the left panel of figure \ref{fig:ISCO_zoom_vert_field_V_70_deg_2}), the estimates provided in tables \ref{tab:Deltas_vert_field} to \ref{tab:Deltas_rad_field_3} are most reliable for low inclinations. 

\subsection{Maximum intensity deviation for equatorial magnetic fields}\label{Subsec::deviation_eq_intensity}

In tables \ref{tab:Deltas_rad_field_1}, \ref{tab:Deltas_rad_field_2} and \ref{tab:Deltas_rad_field_3}, analogously to section \ref{Subsec::deviation_vert_intensity}, we present the maximum intensity deviation along the apparent position of the Kerr analog \ac{ISCO}, $\Delta I$, normalized to the intensity of the Kerr analog $I_\text{Kerr}$ at that point, for equatorial magnetic fields, along with the azimuthal angle $\phi$ at which this maximum occurs.

\begin{table}[htbp]
    \centering
    \footnotesize
    \setlength{\tabcolsep}{8.7pt}
    \renewcommand{\arraystretch}{1.7}

    \begin{tabular}{ccccccc}
        \hline\hline
        \multicolumn{7}{c}{Maximum intensity deviations for $B_r = 0.87$, $B_\theta = 0.0$, $B_\phi = 0.5$}\\
        \hline\hline
        &\multicolumn{3}{c}{Apparent image of the orbit with $R = R^\text{ISCO}_\text{Kerr}$} & \multicolumn{3}{c}{Apparent image of the orbit with $R = \frac{3}{2}R^\text{ISCO}_\text{Kerr}$}\\

        \hline\hline
        
        Model & $i_\text{obs} = 17^\circ$ & $i_\text{obs} = 45^\circ$ &  $i_\text{obs} = 70^\circ$ & $i_\text{obs} = 17^\circ$ & $i_\text{obs} = 45^\circ$ &  $i_\text{obs} = 70^\circ$\\ [1.5ex]
        
        \hline
        
        $\mathbf{I}_{\,0.01}$  & - & -  & - & $\left(-0.35,0.99\pi\right)$ & $\left(-0.43,0.92\pi\right)$ & $\left(-0.56,0.88\pi\right)$\\
        $\mathbf{II}_{\,0.01}$  & $\left(11.2,1.05\pi\right)$ & $\left(6.42,0.98\pi\right)$  & $\left(2.72, 0.99\pi\right)$ &  $\left(-0.09, 0.93\pi\right)$ & $\left(-0.14, 0.89\pi\right)$ & $\left(-0.22, 0.88\pi\right)$ \\
        $\mathbf{III}_{\,0.05}$  & $\left(35.4, 1.00\pi\right)$ & $\left(11.9, 0.89\pi\right)$ & $\left(4.09, 0.83\pi\right)$ & $\left(-0.11, 0.97\pi\right)$ & $\left(-0.33, 0.92\pi\right)$ & $\left(-0.60, 0.89\pi\right)$ \\
        $\mathbf{IV}_{\,0.1}$   & $\left(54.9, 0.99\pi\right)$ & $\left(17.4, 0.89\pi\right)$   & $\left(5.86, 0.83\pi\right)$& $\left(0.67, 0.94\pi\right)$ & $\left(0.45, 0.70\pi\right)$ & $\left(-0.51, 0.90\pi\right)$\\
        $\mathbf{V}_{\,0.2}$    & $\left(54.2, 0.98\pi\right)$ & $\left(20.0, 0.87\pi\right)$   & $\left(7.43, 0.87\pi\right)$& $\left(4.06, 0.94\pi\right)$ & $\left(1.69, 0.85\pi\right)$ & $\left(0.73, 0.79\pi\right)$\\
        $\mathbf{VI}_{\,0.3}$  & $\left(1.73, 0.98\pi\right)$ & $\left(1.10, 0.88\pi\right)$  & $\left(0.66, 0.91\pi\right)$ & $\left(0.63, 0.95\pi\right)$ & $\left(0.36, 0.89\pi\right)$ & $\left(0.20, 0.96\pi\right)$\\
        
        \hline		
        
    \end{tabular}
    
    \caption{\label{tab:Deltas_rad_field_1}\footnotesize The same as table \ref{tab:Deltas_vert_field} but for the magnetic field with cmoponents $B_r = 0.87$, $B_\theta = 0.0$ and $B_\phi = 0.5$.}
    
\end{table}

\begin{table}[htbp]
    \centering
    \footnotesize
    \setlength{\tabcolsep}{8.4pt}
    \renewcommand{\arraystretch}{1.7}
    
    \begin{tabular}{ccccccc}
        \hline\hline
        \multicolumn{7}{c}{Maximum intensity deviations for $B_r = 0.71$, $B_\theta = 0.0$, $B_\phi = 0.71$}\\
        \hline\hline
        &\multicolumn{3}{c}{Apparent image of the orbit with $R = R^\text{ISCO}_\text{Kerr}$} & \multicolumn{3}{c}{Apparent image of the orbit with $R = \frac{3}{2}R^\text{ISCO}_\text{Kerr}$}\\

        \hline\hline
        
        Model & $i_\text{obs} = 17^\circ$ & $i_\text{obs} = 45^\circ$ &  $i_\text{obs} = 70^\circ$ & $i_\text{obs} = 17^\circ$ & $i_\text{obs} = 45^\circ$ &  $i_\text{obs} = 70^\circ$\\ [1.5ex]
        
        \hline
        
        $\mathbf{I}_{\,0.01}$  & - & -  & - & $\left(-0.4, 0.99\pi\right)$ & $\left(-0.46, 0.93\pi\right)$  & $\left(-0.52, 0.88\pi\right)$ \\
        $\mathbf{II}_{\,0.01}$  & $\left(11.1, 1.05\pi\right)$ & $\left(6.66, 0.98\pi\right)$ & $\left(3.01, 0.98\pi\right)$ & $\left(-0.11, 0.95\pi\right)$ & $\left(-0.16, 0.92\pi\right)$ & $\left(-0.17, 0.88\pi\right)$ \\
        $\mathbf{III}_{\,0.05}$  & $\left(34.3, 1.01\pi\right)$ & $\left(10.8, 0.96\pi\right)$ & $\left(8.12, 1.05\pi\right)$ & $\left(-0.55, 0.90\pi\right)$ & $\left(-0.35, 0.93\pi\right)$ & $\left(-0.55, 0.90\pi\right)$ \\
        $\mathbf{IV}_{\,0.1}$   & $\left(50.6, 1.02\pi\right)$ & $\left(14.0, 0.93\pi\right)$ & $\left(11.5, 1.07\pi\right)$ & $\left(0.59, 1.04\pi\right)$ & $\left(0.51, 1.30\pi\right)$ & $\left(-0.55, 0.91\pi\right)$ \\
        $\mathbf{V}_{\,0.2}$    & $\left(47.7, 1.01\pi\right)$ & $\left(15.7, 0.87s\pi\right)$ & $\left(5.19, 0.86\pi\right)$ & $\left(3.58, 0.97\pi\right)$ & $\left(1.58, 0.82\pi\right)$  & $\left(0.88, 0.77\pi\right)$ \\
        $\mathbf{VI}_{\,0.3}$  & $\left(1.65, 1.01\pi\right)$  & $\left(0.95, 0.88\pi\right)$  & $\left(0.51, 0.90\pi\right)$  & $\left(0.59, 0.99\pi\right)$  & $\left(0.35, 0.85\pi\right)$  & $\left(0.15, 0.93\pi\right)$ \\
        
        \hline		
        
    \end{tabular}
    
    \caption{\label{tab:Deltas_rad_field_2}\footnotesize The same as table \ref{tab:Deltas_vert_field} but for the magnetic field with components $B_r = 0.71$, $B_\theta = 0.0$ and $B_\phi = 0.71$.}
    
\end{table}
\begin{table}[htbp]
    \centering
    \footnotesize
    \setlength{\tabcolsep}{8.7pt}
    \renewcommand{\arraystretch}{1.7}
    
    \begin{tabular}{ccccccc}
        \hline\hline
        \multicolumn{7}{c}{Maximum intensity deviations for $B_r = 0.5$, $B_\theta = 0.0$, $B_\phi = 0.87$}\\
        \hline\hline
        &\multicolumn{3}{c}{Apparent image of the orbit with $R = R^\text{ISCO}_\text{Kerr}$} & \multicolumn{3}{c}{Apparent image of the orbit with $R = \frac{3}{2}R^\text{ISCO}_\text{Kerr}$}\\

        \hline\hline
        
        Model & $i_\text{obs} = 17^\circ$ & $i_\text{obs} = 45^\circ$ &  $i_\text{obs} = 70^\circ$ & $i_\text{obs} = 17^\circ$ & $i_\text{obs} = 45^\circ$ &  $i_\text{obs} = 70^\circ$\\ [1.5ex]
        
        \hline
        
        $\mathbf{I}_{\,0.01}$  & - & - & - & $\left(-0.44, 0.99\pi\right)$ & $\left(-0.50, 0.93\pi\right)$ & $\left(-0.49, 0.92\pi\right)$ \\
        $\mathbf{II}_{\,0.01}$  & $\left(11.1, 1.05\pi\right)$ & $\left(6.87, 0.97\pi\right)$ & $\left(2.70, 0.96\pi\right)$ & $\left(-0.14, 0.97\pi\right)$ & $\left(-0.17, 0.93\pi\right)$ & $\left(-0.16, 0.90\pi\right)$ \\
        $\mathbf{III}_{\,0.05}$  & $\left(35.7, 1.04\pi\right)$ & $\left(12.2, 0.98\pi\right)$ & $\left(4.77, 1.01\pi\right)$ & $\left(-0.19, 0.98\pi\right)$ & $\left(-0.38, 0.94\pi\right)$ & $\left(-0.52, 0.93\pi\right)$ \\
        $\mathbf{IV}_{\,0.1}$   & $\left(51.1, 1.03\pi\right)$ & $\left(15.0, 0.97\pi\right)$ & $\left(8.50, 1.03\pi\right)$ & $\left(0.59, 1.10\pi\right)$ & $\left(0.49, 1.27\pi\right)$ & $\left(-0.58, 0.93\pi\right)$ \\
        $\mathbf{V}_{\,0.2}$    & $\left(47.0, 1.03\pi\right)$ & $\left(13.6, 0.89\pi\right)$ & $\left(4.69, 1.01\pi\right)$ & $\left(3.50, 1.01\pi\right)$ & $\left(1.64, 0.79\pi\right)$ & $\left(-0.48, 0.98\pi\right)$ \\
        $\mathbf{VI}_{\,0.3}$  & $\left(1.67, 1.03\pi\right)$ & $\left(0.84, 0.90\pi\right)$ & $\left(0.95, 1.11\pi\right)$ & $\left(0.58, 1.01\pi\right)$ & $\left(0.37, 0.81\pi\right)$ & $\left(0.51, 1.27\pi\right)$ \\
        
        \hline		
        
    \end{tabular}
    
    \caption{\label{tab:Deltas_rad_field_3}\footnotesize The same as table \ref{tab:Deltas_vert_field} but for the magnetic field with components $B_r = 0.5$, $B_\theta = 0.0$ and $B_\phi = 0.87$.}
    
\end{table}

We notice from tables \ref{tab:Deltas_vert_field} - \ref{tab:Deltas_rad_field_3} another unexpected relationship. Models $\mathbf{V}_{0.2}$ and $\mathbf{VI}_{0.3}$ show larger intensity deviations at lower inclinations for the apparent images of both Kerr analog orbits. In contrast models $\mathbf{I}_{0.01}$ - $\mathbf{IV}_{0.1}$ demonstrate this clearly only for $R = R_\text{Kerr}^\text{ISCO}$. This effect holds for both vertical and equatorial fields, which suggests that it is mainly due to the structure of the \ac{TCOs} in the scalarized spacetimes.

\subsection{Maximum EVPA deviations for vertical magnetic fields}\label{Subsec::deviation_vert_evpa}

In this section, analogous to \ref{Subsec::deviation_vert_intensity} and \ref{Subsec::deviation_eq_intensity}, we present the (signed) maximum values of the \ac{EVPA} deviations, across the apparent position of the Kerr analog orbits with radii $R = \{R^\text{ISCO}_\text{Kerr},\frac{3}{2}R^\text{ISCO}_\text{Kerr}\}$, for vertical magnetic fields. Our results are summarized in table \ref{tab:Deltas_vert_field_evpa}. We see the overall trend (which is broken only by model $\mathbf{V}_{0.3}$) that the maximum EVPA deviations increase with the observer's inclination, and decrease radially outwards along the image. \bigskip\par

Unlike the intensity deviations $\Delta I$, the EVPA deviations $\Delta \text{EVPA}$ are bounded in the interval $[-\frac{\pi}{2}, \frac{\pi}{2}]$. The bounding values $\pm\frac{\pi}{2}$ correspond to the polarization vectors appearing orthogonal when projected on the observer's screen. When this occurs for a given image, we label it in table \ref{tab:Deltas_vert_field_evpa} as $(\frac{\max \Delta \text{EVPA}}{\pi},\,\frac{\phi}{\pi}) = (\pm0.5,\frac{\phi}{\pi})$. We notice that the only models that show an \ac{EVPA} deviation of $\pm\frac{\pi}{2}$ are the least scalarized ones $\mathbf{V}_{0.2}$ and $\mathbf{VI}_{0.3}$.\bigskip\par

\newpage

Another interesting feature is the sign changes when increasing the observer inclination (most readily seen for models $\mathbf{III}_{0.05}$ and $\mathbf{IV}_{0.1}$). This is due to the graph of $\Delta\text{EVPA}$ having more than one extrema under certain conditions (similar to figures \ref{fig:ISCO_zoom_vert_field_II_70_deg}, \ref{fig:2.5_ISCO_zoom_vert_field_V_70_deg} and \ref{fig:3.5_ISCO_zoom_vert_field_II_70_deg}). As the observer inclination varies, the relative size of those extrema changes at different rates. This results at the maximum \ac{EVPA} deviation changing sign as observer inclination varies.

\begin{table}[htbp]
    \centering
    \footnotesize
    \setlength{\tabcolsep}{10.7pt}
    \renewcommand{\arraystretch}{1.7}
    
    \begin{tabular}{ccccccc}
        \hline\hline
        \multicolumn{7}{c}{Maximum EVPA deviations for $B_r = 0.0$, $B_\theta = 1.0$, $B_\phi = 0.0$}\\
        \hline\hline
        &\multicolumn{3}{c}{Apparent image of the orbit with $R = R^\text{ISCO}_\text{Kerr}$} & \multicolumn{3}{c}{Apparent image of the orbit with $R = \frac{3}{2}R^\text{ISCO}_\text{Kerr}$}\\

        \hline
        Model & $i_\text{obs} = 17^\circ$ & $i_\text{obs} = 45^\circ$ &  $i_\text{obs} = 70^\circ$ & $i_\text{obs} = 17^\circ$ & $i_\text{obs} = 45^\circ$ &  $i_\text{obs} = 70^\circ$\\ [1.5ex]
        
        \hline
        
        $\mathbf{I}_{\,0.01}$  & - & -  & - & $\left(0.008, 0.41\right)$ & $\left(0.01, 0.39\right)$  & $\left(0.04, 0.85\right)$ \\
        $\mathbf{II}_{\,0.01}$  & $\left(0.01,0
        27\right)$ & $\left(0.01,0.37\right)$ & $\left(-0.05,0.85\right)$ & $\left(0.004,0.25\right)$ & $\left(0.005,0.3\right)$ & $\left(0.02, 0.84\right)$ \\
        $\mathbf{III}_{\,0.05}$  & $\left(0.03, 0.28\right)$ & $\left(-0.09,0.84\right)$ & $\left(-0.27, 0.77\right)$ & $\left(0.01,0.36\right)$ & $\left(0.02,0.44\right)$ & $\left(0.03, 0.48\right)$ \\
        $\mathbf{IV}_{\,0.1}$   & $\left(0.06,0.33\right)$ & $\left(-0.13,0.85\right)$ & $\left(-0.38, 0.77\right)$ & $\left(0.03,0.36\right)$ & $\left(-0.06, 0.88\right)$ & $\left(-0.21, 0.78\right)$ \\
        $\mathbf{V}_{\,0.2}$    & $\left(0.16, 0.40\right)$ & $\left(0.32, 0.59\right)$ & $\left(\pm0.5, 0.52\right)$ & $\left(0.15, 0.47\right)$ & $\left(\pm0.5, 0.81\right)$  & $\left(\pm0.5, 0.52\right)$ \\
        $\mathbf{VI}_{\,0.3}$  & $\left(0.06, 0.4\right)$  & $\left(-0.13, 0.84\right)$  & $\left(\pm0.5,0.77\right)$  & $\left(0.07, 0.52\right)$  & $\left(0.49, 0.69\right)$  & $\left(0.27, 0.60\right)$ \\
        
        \hline		
        
    \end{tabular}
    
    \caption{\label{tab:Deltas_vert_field_evpa}\footnotesize The maximum (signed) EVPA deviation $\Delta \text{EVPA}$ along the apparent position of two Kerr analog orbits (with radii $R = \{R^\text{ISCO}_\text{Kerr}, \frac{3}{2}R^\text{ISCO}_\text{Kerr}\}$) for all six solutions considered in this paper, for a vertical magnetic field. The reported values are presented as $(\frac{\max \Delta \text{EVPA}}{\pi},\,\frac{\phi}{\pi})$, where $\phi$ is the azimuthal angle along the image at which the maxima occur. For the specific case R = $R^\text{ISCO}_\text{Kerr}$, the apparent position of this orbit does not entirely overlap with the image of the disk for model $\mathbf{I}_{0.01}$. We therefore do not specify maximum EVPA deviations for this case.}
    
\end{table}

\subsection{Maximum EVPA deviations for equatorial magnetic fields}\label{Subsec::deviation_eq_evpa}

In this section we present the maximum \ac{EVPA} deviations for the three equatorial fields considered in this paper. The chosen orbits are again those with radii $R = \{R^\text{ISCO}_\text{Kerr}, \frac{3}{2}R^\text{ISCO}_\text{Kerr}\}$, and our results are presented in tables \ref{tab:Deltas_eq_field_evpa_1}, \ref{tab:Deltas_eq_field_evpa_2} and \ref{tab:Deltas_eq_field_evpa_3}. For the magnetic field with components $\vec{B} = (0.87, 0.0, 0.5)$ we find the expected behavior -- $\max \Delta\text{EVPA}$ decreases in amplitude as the orbit radius increases, and increases in amplitude as the inclination increases. For the other two magnetic field configurations, this does not always hold true. The more highly scalarized models $\mathbf{II}_{0.01}$ and $\mathbf{III}_{0.05}$ demonstrate a higher amplitude for $\Delta\text{EVPA}$ at the higher orbit $R = \frac{3}{2}R^\text{ISCO}_\text{Kerr}$.\bigskip\par

It is also interesting to note that no equatorial magnetic field produces deviations in the \ac{EVPA} as large as those seen in the vertical field case (table \ref{tab:Deltas_vert_field_evpa}). This is a property of the magnetic field (and the polarization pattern it produces) rather than a consequence of the presence of a massive scalar field. Figure \ref{fig:Whole_disk_vert_field_II_V_70_deg} shows that in the high inclincation case, a vertical magnetic field produces a characteristic cusp-like structure of the polarization pattern near the upper left edge of the shadow, varies rapidly as one goes radially out along the image. The combination of this pattern with the large difference in the radial source coordinate along the images we consider (the left panel of figure \ref{fig:ISCO_zoom_vert_field_V_70_deg_2}) creates a significant deviation in the \ac{EVPA}. In contrast, equatorial magnetic fields produce a polarization pattern that does not vary as quickly along the image of the disk. This naturally produces smaller overall deviations in the \ac{EVPA}.

\begin{table}[htbp]
    \centering
    \footnotesize
    \setlength{\tabcolsep}{10.7pt}
    \renewcommand{\arraystretch}{1.7}
    
    \begin{tabular}{ccccccc}
        \hline\hline
        \multicolumn{7}{c}{Maximum EVPA deviations for $B_r = 0.87$, $B_\theta = 0.0$, $B_\phi = 0.5$}\\
        \hline\hline
        &\multicolumn{3}{c}{Apparent image of the orbit with $R = R^\text{ISCO}_\text{Kerr}$} & \multicolumn{3}{c}{Apparent image of the orbit with $R = \frac{3}{2}R^\text{ISCO}_\text{Kerr}$}\\

        \hline
        Model & $i_\text{obs} = 17^\circ$ & $i_\text{obs} = 45^\circ$ &  $i_\text{obs} = 70^\circ$ & $i_\text{obs} = 17^\circ$ & $i_\text{obs} = 45^\circ$ &  $i_\text{obs} = 70^\circ$\\ [1.5ex]
        
        \hline
        
        $\mathbf{I}_{\,0.01}$  & - & -  & - & $\left(-0.02,0.88\right)$ & $\left(-0.05, 0.87\right)$  & $\left(-0.14, 0.92\right)$ \\
        $\mathbf{II}_{\,0.01}$  & $\left(0.04,0.59\right)$ & $\left(0.05,0.64\right)$ & $\left(0.07,0.69\right)$ & $\left(-0.01, 0.84\right)$ & $\left(-0.02,0.87\right)$ & $\left(-0.06,0.92\right)$ \\
        $\mathbf{III}_{\,0.05}$ & $\left(0.12,0.72\right)$ & $\left(0.13,0.85\right)$ &  $\left(0.18,0.9\right)$ & $\left(-0.006,0.89\right)$  & $\left(-0.03,0.87\right)$ & $\left(-0.12, 0.91\right)$ \\
        $\mathbf{IV}_{\,0.1}$ & $\left(0.12, 0.84\right)$ & $\left(0.14,0.94\right)$ & $\left(0.22,0.93\right)$ & $\left(0.025,1.07\right)$ & $\left(0.03,1.05\right)$ & $\left(0.04,0.99\right)$ \\
        $\mathbf{V}_{\,0.2}$ & $\left(0.11, 1.05\right)$ & $\left(0.16, 0.94\right)$ & $\left(0.34,0.96\right)$ & $\left(0.07, 1.17\right)$ & $\left(0.11,1.07\right)$  & $\left(0.26,1.01\right)$ \\
        $\mathbf{VI}_{\,0.3}$ & $\left(0.03, 1.07\right)$  & $\left(0.05, 1.00\right)$  & $\left(0.11,0.96\right)$  & $\left(0.02, 1.25\right)$  & $\left(0.03, 1.10\right)$ & $\left(0.08,1.06\right)$ \\
        
        \hline		
        
    \end{tabular}
    
    \caption{\label{tab:Deltas_eq_field_evpa_1}\footnotesize The same as table \ref{tab:Deltas_vert_field_evpa} but for the magnetic field with components $B_r = 0.87$, $B_\theta = 0.0$ and $B_\phi = 0.5$.}
    
\end{table}

\begin{table}[htbp]
    \centering
    \footnotesize
    \setlength{\tabcolsep}{10.1pt}
    \renewcommand{\arraystretch}{1.7}
    
    \begin{tabular}{ccccccc}
        \hline\hline
        \multicolumn{7}{c}{Maximum EVPA deviations for $B_r = 0.71$, $B_\theta = 0.0$, $B_\phi = 0.71$}\\
        \hline\hline
        &\multicolumn{3}{c}{Apparent image of the orbit with $R = R^\text{ISCO}_\text{Kerr}$} & \multicolumn{3}{c}{Apparent image of the orbit with $R = \frac{3}{2}R^\text{ISCO}_\text{Kerr}$}\\

        \hline
        Model & $i_\text{obs} = 17^\circ$ & $i_\text{obs} = 45^\circ$ &  $i_\text{obs} = 70^\circ$ & $i_\text{obs} = 17^\circ$ & $i_\text{obs} = 45^\circ$ &  $i_\text{obs} = 70^\circ$\\ [1.5ex]
        
        \hline
        
        $\mathbf{I}_{\,0.01}$  & - & -  & - & $\left(-0.01, 0.86\right)$ & $\left(-0.03, 0.85\right)$  & $\left(-0.11, 0.89\right)$ \\
        $\mathbf{II}_{\,0.01}$  & $\left(0.02, 0.56\right)$ & $\left(0.03, 0.59\right)$ & $\left(0.04, 0.59\right)$ & $\left(-0.007, 0.84\right)$ & $\left(-0.02, 0.85\right)$ & $\left(-0.05, 0.90\right)$ \\
        $\mathbf{III}_{\,0.05}$ & $\left(0.06, 0.68\right)$ & $\left(0.07, 0.54\right)$ & $\left(0.09, 0.89\right)$ & $\left(-0.007, 0.88\right)$ & $\left(-0.03, 0.86\right)$ & $\left(-0.098, 0.89\right)$ \\
        $\mathbf{IV}_{\,0.1}$ & $\left(0.06, 0.76\right)$ & $\left(0.07, 0.93\right)$ & $\left(0.12, 0.92\right)$ & $\left(0.01, 1.12\right)$ & $\left(0.01, 1.06\right)$ & $\left(-0.04, 0.85\right)$ \\
        $\mathbf{V}_{\,0.2}$ & $\left(0.06, 1.07\right)$ & $\left(0.10, 0.93\right)$ & $\left(0.27, 0.93\right)$ & $\left(0.04, 1.14\right)$ & $\left(0.08, 1.01\right)$  & $\left(0.20, 0.98\right)$ \\
        $\mathbf{VI}_{\,0.3}$ & $\left(0.02, 1.07\right)$  & $\left(0.04, 0.94\right)$  & $\left(0.12, 0.96\right)$  & $\left(0.01, 1.16\right)$  & $\left(0.03, 1.03\right)$ & $\left(0.08, 1.01\right)$ \\
        
        \hline		
        
    \end{tabular}
    
    \caption{\label{tab:Deltas_eq_field_evpa_2}\footnotesize The same as table \ref{tab:Deltas_vert_field_evpa} but for the magnetic field with components $B_r = 0.7$1, $B_\theta = 0.0$ and $B_\phi = 0.71$.}
    
\end{table}

\begin{table}[htbp]
    \centering
    \footnotesize
    \setlength{\tabcolsep}{9.2pt}
    \renewcommand{\arraystretch}{1.7}
    
    \begin{tabular}{ccccccc}
        \hline\hline
        \multicolumn{7}{c}{Maximum EVPA deviations for $B_r = 0.5$, $B_\theta = 0.0$, $B_\phi = 0.87$}\\
        \hline\hline
        &\multicolumn{3}{c}{Apparent image of the orbit with $R = R^\text{ISCO}_\text{Kerr}$} & \multicolumn{3}{c}{Apparent image of the orbit with $R = \frac{3}{2}R^\text{ISCO}_\text{Kerr}$}\\

        \hline
        Model & $i_\text{obs} = 17^\circ$ & $i_\text{obs} = 45^\circ$ &  $i_\text{obs} = 70^\circ$ & $i_\text{obs} = 17^\circ$ & $i_\text{obs} = 45^\circ$ &  $i_\text{obs} = 70^\circ$\\ [1.5ex]
        
        \hline
        
        $\mathbf{I}_{\,0.01}$  & - & -  & - & $\left(-0.01, 0.86\right)$ & $\left(-0.03, 0.85\right)$  & $\left(-0.09, 0.87\right)$ \\
        $\mathbf{II}_{\,0.01}$  & $\left(0.008, 0.42\right)$ & $\left(0.01, 0.43\right)$ & $\left(0.014, 0.47\right)$ & $\left(-0.006, 0.84\right)$ & $\left(-0.01, 0.84\right)$ & $\left(-0.04, 0.88\right)$ \\
        $\mathbf{III}_{\,0.05}$ & $\left(0.02, 0.41\right)$ & $\left(0.021, 0.38\right)$ & $\left(0.02, 0.39\right)$ & $\left(-0.01, 0.84\right)$ & $\left(-0.026, 0.84\right)$ & $\left(-0.08, 0.88\right)$ \\
        $\mathbf{IV}_{\,0.1}$ & $\left(0.01, 0.29\right)$ & $\left(0.015, 0.01\right)$ & $\left(0.03, 0.92\right)$ & $\left(-0.006, 0.61\right)$ & $\left(-0.013, 0.73\right)$ & $\left(-0.04, 0.85\right)$ \\
        $\mathbf{V}_{\,0.2}$ & $\left(0.01, 1.15\right)$ & $\left(0.03, 0.88\right)$ & $\left(0.20, 0.89\right)$ & $\left(-0.02, 0.29\right)$ & $\left(-0.05, 0.31\right)$  & $\left(0.15, 0.95\right)$ \\
        $\mathbf{VI}_{\,0.3}$ & $\left(-0.006, 0.27\right)$  & $\left(0.02, 0.89\right)$  & $\left(0.10, 0.95\right)$  & $\left(-0.01, 0.28\right)$  & $\left(-0.03, 0.29\right)$ & $\left(-0.08, 0.27\right)$ \\
        
        \hline		
        
    \end{tabular}
    
    \caption{\label{tab:Deltas_eq_field_evpa_3}\footnotesize The same as table \ref{tab:Deltas_vert_field_evpa} but for the magnetic field with components $B_r = 0.5$, $B_\theta = 0.0$ and $B_\phi = 0.87$.}
    
\end{table}

\newpage

\section{Conclusion}\label{Sec:Conclustions}

In this work, we have investigated the polarization properties of direct images of geometrically and optically thin accretion disks, emitting synchrotron radiation around Kerr black holes with synchronized scalar hair. We model the emission process with a simple analytical prescription of a fluid ring in the equatorial plane of the black hole, which nevertheless captures well the phenomenology of synchrotron emission at the observational frequency of the \ac{EHT}.  We then numerically solve the geodesic equations of motion for the light rays, starting from the observer and ending at the emission point on the accretion disk. The emission model then specifies a polarization vector at that point, which we numerically parallel transport to the observer to form the final image. For each considered scalarized solution we define an analogous Kerr black hole in GR with the same value of the ADM mass and spin parameter, to which we compare the polarization properties of the images.\bigskip\par

A key result of this work is that at low inclinations the main difference between the polarization patterns of the scalarized solutions and their corresponding Kerr analogs (that is independent of the radial flux distribution) is a dephasing in the \ac{EVPA} pattern. This effect can partially be attributed to the fact that the scalar field in these solutions forms a torus around the central black hole, which extends to significantly larger radii than the event horizon, and contains a significant portion of the angular momentum. Therefore, photons moving through the scalar field torus experience a different frame-dragging effect compared to the Kerr analog. This influences not only the process of parallel transport of the polarization vector, but also the azimuthal position of the source for fixed coordinates on the image.\bigskip\par

At a higher inclination of $70^\circ$ we find that images formed in a vertical magnetic field show a characteristic reversal in the twist direction of the polarization vector. This effect occurs for both the scalarized solutions and their Kerr analogs, for orbits sufficiently far away from the \ac{ISCO}. We find that across the apparent position of the Kerr analog \ac{ISCO}, solutions $\mathbf{V}_{0.2}$ and $\mathbf{V}_{0.3}$ present this reversal, while the rest do not. Looking at higher orbital radii, the effect becomes apparent in all scalarized solutions. We attribute this behavior to the different focusing properties of the solutions, which result from the distribution of mass between the black hole and scalar field. Images formed with equatorial magnetic fields do not exhibit this reversal of the \ac{EVPA} for any of the scalarized solutions considered in this paper.\bigskip\par

A particularly important and somewhat unexpected result is that the deviations in the polarization pattern, as measured by the EVPA, are more pronounced for models with smaller normalized Noether charge $q$. This is contrary to the black hole shadows  which deviate from Kerr only moderately for smaller $q$ \cite{cunha2016shadows,Cunha2016chaotic_lensing,Galin_2026}. This indicates that the polarization signal is not solely determined by the total amount of scalar hair, but is instead sensitive to how the scalar field modifies the local spacetime structure in the emission region and along photon trajectories. While larger values of $q$ correspond to stronger global deviations from the Kerr geometry, they also imply that a larger fraction of the mass and angular momentum is carried by the scalar field. In contrast, models with small normalized Noether charge $q$ remain closer to Kerr globally, but introduce localized modifications in the near-horizon region and in the geodesic structure that governs photon propagation. In particular, the position of the \ac{ISCO} orbits shifts, depending on the distribution of mass between the central black hole and the scalar cloud. This can cause photons emitted between the black hole and the scalar cloud to deviate substantially from their Kerr analogs. Since the observed polarization is determined by the parallel transport of the polarization vector along these geodesics, such localized differences can lead to a larger accumulated phase shift and, consequently, stronger EVPA dephasing.\bigskip\par

These results demonstrate that polarization observables provide a sensitive probe of local variations of the spacetime geometry rather than global properties alone, and are therefore particularly sensitive to intermediate or weak deviations from the Kerr solution. This highlights the strong potential of polarized imaging as a diagnostic tool for detecting and constraining scalar hair in astrophysical black hole environments. Future high-resolution polarimetric observations are expected to provide a complementary probe to shadow measurements and accretion disk emission in testing deviations from the Kerr paradigm.

\begin{appendix}

\section{Physical quantities of the considered solutions}\label{Appendix Solutons}

Below in table \ref{tab_sol_properties} we provide a detailed description of the physical properties of the solutions, considered in this paper. They are taken from \cite{collodel2020rotating} in the case of a flat target space metric. The labels in the first column coincide with the labels in figure \ref{fig:M-Omega Space}.

\begin{table}[H]
  \centering
  \footnotesize
    \setlength{\tabcolsep}{7.7pt} 
    \renewcommand{\arraystretch}{1.4} 
    \vspace{1.5mm} 
  \begin{tabular}{lccccccccccc}
    \hline\hline
    Label & $\omega_{s}/\mu$ & $M\mu$ & $J\mu^2$ & $M_{BH}\mu$ & $J_{BH}\mu^2$ & $M_{\psi}\mu$ & $J_{\psi}\mu^2$ & $J_{\psi}/J$ & $J/M^2$ & $J_{BH}/M_{BH}^2$ \\
    \hline
    $\mathbf{I}_{\,0.01}$  & 0.6792 & 0.8820 & 0.7259 & 0.0035 & 0.0001 & 0.8785 & 0.7258 & 0.9999 & 0.9331 & 5.8772 \\
    $\mathbf{II}_{\,0.01}$   & 0.8353 & 0.6482 & 0.4068 & 0.0049 & 0.0011 & 0.6434 & 0.4057 & 0.9973 & 0.9680 & 46.444 \\    
    $\mathbf{III}_{\,0.05}$ & 0.7064 & 0.9082 & 0.8329 & 0.0252 & 0.0045 & 0.9415 & 0.8283 & 0.9945 & 0.8915 & 7.1310 \\
    $\mathbf{IV}_{\,0.1}$   & 0.7385 & 1.0004 & 0.8535 & 0.0870 & 0.0302 & 0.9134 & 0.8232 & 0.9645 & 0.8528 & 3.9982 \\
    $\mathbf{V}_{\,0.2}$    & 0.8955 & 0.8787 & 0.6790 & 0.2796 & 0.1012 & 0.5990 & 0.5776 & 0.8508 & 0.8793 & 1.2946 \\
    $\mathbf{VI}_{\,0.3}$   & 0.9880 & 0.3200 & 0.1260 & 0.2377 & 0.0444 & 0.0823 & 0.0816 & 0.6478 & 1.2304 & 0.7856 \\
    \hline\hline
  \end{tabular}
  \caption{\label{tab_sol_properties}\footnotesize Physical quantities of hairy black hole solutions, corresponding to a zero Gaussian curvature of the target space, $\kappa=0$. Each model is labeled as $\mathbf{X}_{\,v}$, where $\mathbf{X}$ denotes the model number, and the subscript $v$ indicates the black hole horizon radius $r_{\rm H}$.}   
\end{table}

Table \ref{tab:Kerr_analog_spins} shows the normalized spin parameters of the Kerr analogs. These spin parameters are chosen so that the black hole has the same ADM mass and horizon radius, equal to that of the scalarized model. This results in an expression for $a_\textbf{Kerr}$, given by eq. \ref{Kerr_spin}. 

\begin{table}[htbp]
    \centering
    \small
    \setlength{\tabcolsep}{4pt}
    \renewcommand{\arraystretch}{1.7}

    \begin{tabular}{ccccccc}
        \hline\hline
        \multicolumn{7}{c}{Kerr analog spin parameters}\\
        \hline\hline
        Model & $\mathbf{I}_{0.01}$ & $\mathbf{II}_{0.01}$ & $\mathbf{III}_{0.05}$ & $\mathbf{IV}_{0.1}$ & $\mathbf{V}_{0.2}$ & $\mathbf{VI}_{0.3}$\\
        $a_\text{Kerr}/M_\text{ADM}$ & 0.999988 & 0.999971 &  0.999664 & 0.998750 & 0.993506 &  0.883338\\ [1.5ex]
        
        \hline		
        
    \end{tabular}
    
    \caption{\centering\label{tab:Kerr_analog_spins}\footnotesize The Kerr analog spin parameter for each considered numerical model.}
    
\end{table}

\section{\ac{ISCO} structure of the considered solutions}

In table \ref{tab:ProgradeTCOs2} we present the locations of the \ac{ISCO} for the scalarized solutions, as well as their Kerr analogs. We give the Kerr analog orbit radii in both Boyer-Lindquist coordinates and the coordinates of line element \eqref{lucas line element}. The conversion between them is given by:
\begin{equation}
    r = R - \frac{a^2}{R_H},
\end{equation}

\noindent where the spin parameter of the Kerr analog is given by \eqref{Kerr_spin}, $r$ is the radial coordinate of line element \eqref{lucas line element} and $R$ is the Boyer-Lindquist radial coordinate. We note that model $\mathbf{I}_{0.01}$ has two disjoint prograde regions, and thus two prograde \ac{ISCO}. In this case, we consider the disk as starting from the outer \ac{ISCO}. The other five models have only one \ac{ISCO}. A more thorough discussion on the structure of these orbits can be found in \cite{Galin_2026}.

\begin{table}[htbp]
    \centering
    \footnotesize
    \setlength{\tabcolsep}{24.1pt}
    \renewcommand{\arraystretch}{1.3}
    
    \begin{tabular}{c*{4}{c}}
        \hline\hline
        Model & $R^{\text{ISCO}}_\text{Kerr}$ & $r^{\text{ISCO}}_{\text{Kerr}}$ &  $r_{+}^{\text{ISCO}}$ \\
        \hline
        $\mathbf{I}_{\,0.01}$  & $0.9154$ & $0.0401$  & 0.2685\\
        $\mathbf{II}_{\,0.01}$  & $0.6810$ & $0.0377$  & 0.0539 \\
        $\mathbf{III}_{\,0.05}$  & $1.0839$ & $0.1424$   & 0.2454 \\
        $\mathbf{IV}_{\,0.1}$   & $1.1692$ & $0.2188$   & 0.2403 \\
        $\mathbf{V}_{\,0.2}$    & $1.2121$ & $0.4334$   & 0.402 \\
        $\mathbf{VI}_{\,0.3}$  & $0.7776$ & $0.6076$   & 0.6330 \\
        \hline		
    \end{tabular}
    
    \caption{\label{tab:ProgradeTCOs2}\footnotesize The locations of the prograde \ac{ISCO} orbits for models $\mathbf{I}_{\,0.01}$\,--\,$\mathbf{VI}_{\,0.3}$ and their Kerr analogs. We give the Kerr analog \ac{ISCO} in both Boyer-Lindquist coordinates, labeled as $R^\textbf{ISCO}_\text{Kerr}$, and the coordinates of line element \eqref{lucas line element}, labeled $r^\text{ISCO}_\text{Kerr}$. See \cite{Galin_2026} for a detailed discussion regarding the \ac{TCOs} and light ring structure of the scalarized models.}
\end{table}
\end{appendix}

\section*{Acknowledgments}

This study is financed by the Bulgarian National Science Fund (NSF) under Grant KP-06-DV/8 within the funding programme ``VIHREN--2024''. D.D. acknowledges financial support from the Spanish Ministry of Science and Innovation through the Ram\'on y Cajal programme (grant RYC2023-042559-I), funded by MCIN/AEI/\href{https://doi.org/10.13039/501100011033}{10.13039/50110001\-1033}, from an Emmy Noether Research Group funded by the German Research Foundation (DFG) under Grant No.~DO~1771/1-1, and by the Spanish Agencia Estatal de Investigaci\'on (grant PID2024-159689NB-C21) funded by MICIU/AEI/10.13039/501100011033 and by FEDER / EU. S.Y. is supported by the European Union-NextGenerationEU, through the National Recovery and Resilience Plan of the Republic of Bulgaria, project No. BG-RRP-2.004-0008-C01.

\bibliography{References}

\end{document}